\documentclass[sigconf]{acmart}

\usepackage{enumitem}
\usepackage[captionskip=-1pt]{subfig}
\usepackage{graphicx}
\usepackage{multirow}
\usepackage{makecell}
\usepackage{algorithm,algorithmic}
\usepackage{pifont}

\theoremstyle{plain}

\allowdisplaybreaks

\AtBeginDocument{%
  }

\copyrightyear{2025} 
\acmYear{2025} 
\setcopyright{cc}
\setcctype{by}
\acmConference[WWW '25]{Proceedings of the ACM Web Conference 2025}{April
28-May 2, 2025}{Sydney, NSW, Australia}
\acmBooktitle{Proceedings of the ACM Web Conference 2025 (WWW '25), April
28-May 2, 2025, Sydney, NSW, Australia}
\acmDOI{10.1145/3696410.3714760}
\acmISBN{979-8-4007-1274-6/25/04}

\begin{document}

\title{Polynomial Selection in Spectral Graph Neural Networks: An Error-Sum of Function Slices Approach}

\author{Guoming Li}
\authornote{Corresponding Authors}
\affiliation{%
  \institution{Mohamed bin Zayed University of Artificial Intelligence}
  \city{Abu Dhabi}
  \country{United Arab Emirates}
}
\email{paskardli@outlook.com}

\author{Jian Yang}
\affiliation{%
  \institution{University of Chinese Academy of Sciences}
  \city{Beijing}
  \country{China}
}
\email{jianyang0227@gmail.com}

\author{Shangsong Liang}
\affiliation{%
  \institution{Sun Yat-sen University}
  \city{Guangzhou}
  \country{China}
}
\email{liangshangsong@gmail.com}

\author{Dongsheng Luo}
\authornotemark[1]
\affiliation{%
  \institution{Florida International University}
  \city{Miami}
  \country{United States}
}
\email{dluo@fiu.edu}

\renewcommand{\shortauthors}{Li et al.}

\begin{abstract}
Spectral graph neural networks are proposed to harness spectral information inherent in graph-structured data through the application of polynomial-defined graph filters, recently achieving notable success in graph-based web applications. 
Existing studies reveal that various polynomial choices greatly impact spectral GNN performance, underscoring the importance of polynomial selection. 
However, this selection process remains a critical and unresolved challenge. 
Although prior work suggests a connection between the approximation capabilities of polynomials and the efficacy of spectral GNNs, there is a lack of theoretical insights into this relationship, rendering polynomial selection a largely heuristic process.

To address the issue, this paper examines polynomial selection from an error-sum of function slices perspective. 
Inspired by the conventional signal decomposition, we represent graph filters as a sum of disjoint function slices. 
Building on this, we then bridge the polynomial capability and spectral GNN efficacy by proving that the construction error of graph convolution layer is bounded by the sum of polynomial approximation errors on function slices. 
This result leads us to develop an advanced filter based on trigonometric polynomials, a widely adopted option for approximating narrow signal slices. 
The proposed filter remains provable parameter efficiency, with a novel Taylor-based parameter decomposition that achieves streamlined, effective implementation. 
With this foundation, we propose TFGNN, a scalable spectral GNN operating in a decoupled paradigm. 
We validate the efficacy of TFGNN via benchmark node classification tasks, along with an example graph anomaly detection application to show its practical utility.
\end{abstract}




\received{20 February 2007}
\received[revised]{12 March 2009}
\received[accepted]{5 June 2009}

\maketitle


\section{Introduction}
Graph neural networks (GNNs)~\cite{comprehensivegnn,gnn-survey} have emerged as powerful tools to capture structural information from graph data, 
facilitating advanced performance across numerous web applications, such as web search~\cite{websearch_1,websearch_2}, recommender system~\cite{recsys_1,recsys_2}, social network analysis~\cite{social_2}, anomaly detection~\cite{GAnoDet-1-CARE-GNN,GAnoDet-2-PC-GNN,GAnoDet-3-GDN}, etc.
Among GNN varieties, spectral GNNs stand out for their ability to exploit the spectral properties of graph data using polynomial-defined graph filters, recently achieving notable success in graph-related tasks~\cite{surveyspectralgnn}.

Numerous existing studies have empirically revealed that various polynomial choices greatly impact spectral GNN performance~\cite{BernNet-GNN-narrowbandresults-1,ChebNetII,JacobiConv,OptBasisGNN,decoupled-PCConv,decoupled-UniFilter}, underscoring the importance of polynomial selection. 
However, despite the various works that incorporate different polynomials, their primary focus has been on other factors, such as convergence rate~\cite{OptBasisGNN,JacobiConv}, rather than explicitly targeting the enhancement of spectral GNN efficacy. 
As far as we are aware, there is no existing work that directly associates spectral GNN efficacy with polynomial capability, which renders polynomial selection a crucial yet unresolved challenge, often approached heuristically.

To tackle this issue, we investigate polynomial selection through a novel lens of error-sum of function slices in this paper. 
Drawing inspiration from signal decomposition techniques~\cite{Digitalsignalprocessing}, we uniformly represent graph filters as a sum of disjoint function slices. 
We present the first proof establishing that the construction error of graph convolution layers is bounded by the sum of polynomial approximation errors on these function slices. 
This explicitly links the capability of polynomials to the effectiveness of spectral GNNs, supported by intuitive numerical validations that affirm the practicality of our theoretical framework. 
This finding emphasizes that enhanced spectral GNN efficacy can be attained by utilizing graph filters created with ``narrow slice-preferred polynomials''. 
Consequently, we introduce an innovative filter based on trigonometric polynomials~\cite{TrigonometricSeries}, a standard approach for approximating narrow signal slices in the signal processing domain. 
Our proposed filter showcases proven parameter efficiency, leveraging a novel Taylor-based parameter decomposition that facilitates streamlined and effective implementation. 
Building upon this foundation, we introduce TFGNN, a scalable spectral GNN operating in a widely adopted decoupled GNN architecture~\cite{decoupled-advantages-3-coupled-disadvantages-2-APPNP,GPRGNN,BernNet-GNN-narrowbandresults-1,decoupled-NFGNN,decoupled-PCConv}. 
Empirically, we validate TFGNN’s capacity via benchmark node classification tasks and highlight its real-world efficacy with an example graph anomaly detection application. 
Our contributions are summarized below:
\begin{itemize}[leftmargin=15pt,parsep=0pt,itemsep=2pt,topsep=2pt]
\item We provide the inaugural proof that connects the efficacy of spectral GNN to their polynomial capabilities, framed through the lens of approximation error on function slices. Our numerical experiments reinforce the practical utility of this connection. This finding offers an informed strategy to refine polynomial selection, leading to enhanced spectral GNNs.
\item We introduce an advanced graph filter based on trigonometric polynomials, showcasing provable parameter efficiency. Our novel approach incorporates a Taylor-based parameter decomposition to achieve a streamlined implementation. Based on this filter, we further develop TFGNN, a scalable spectral GNN characterized by its decoupled architecture.
\item We validate TFGNN's effectiveness with extensive experiments in benchmark node classification and an illustrative application in graph anomaly detection. The results reveal that TFGNN not only exceeds previous methods in standard tasks but also yields results comparable to specialized models in real-world settings, demonstrating its significant practical value.
\end{itemize}


\section{Backgrounds and Preliminaries}
\label{section-preliminaries}

\subsubsection*{\bf Graph notations} Let $\mathcal{G}=(\boldsymbol{A}, \boldsymbol{X})$ be an undirected and unweighted graph with adjacency matrix $\boldsymbol{A}\in\{0,1\}^{n\times n}$ and node feature $\boldsymbol{X}\in\mathbb{R}^{n\times m}$. 
In addition, $\boldsymbol{L}=\boldsymbol{I}-\boldsymbol{D}^{-\frac{1}{2}}\boldsymbol{A}\boldsymbol{D}^{-\frac{1}{2}}$ is the \textit{normalized graph Laplacian}~\cite{spectralgraphtheory}, with $\boldsymbol{I}$, $\boldsymbol{D}$ being the identity matrix and the degree matrix, respectively. 
The eigen-decomposition of $\boldsymbol{L}$ is given by $\boldsymbol{L}=\boldsymbol{U}diag(\boldsymbol{\lambda})\boldsymbol{U}^{T}$, where $\boldsymbol{U}\in\mathbb{R}^{n\times n}$ denotes the eigenvectors, and $\boldsymbol{\lambda}\in\left[0,2\right]^{n}$ represents the corresponding eigenvalues.

\subsubsection*{\bf Graph filters} The concept of graph filters originates in the field of Graph Signal Processing (GSP)~\cite{GraphSignalProcessingOverviewChallengesandApplications,DiscreteSignalProcessingonGraphs,Theemergingfield}, a field dedicated to developing specialized tools for processing signals generated on graphs, grounded in spectral graph theory~\cite{spectralgraphtheory}. 
A graph filter is specifically a point-wise mapping $f: \left[0,2\right] \mapsto \mathbb{R}$ applied to graph Laplacian's eigenvalues, $\boldsymbol{\lambda}$, facilitating the processing of the graph signal $\boldsymbol{x}\in\mathbb{R}^{n}$ through a filtering operation as shown below~\cite{DiscretesignalprocessingongraphsGraphfilters}:
\begin{align}
\boldsymbol{z} \triangleq \boldsymbol{U}diag(f(\boldsymbol{\lambda}))\boldsymbol{U}^{T}\boldsymbol{x}\ ,
\label{eq:graph-filtering}
\end{align}
where $\boldsymbol{z}\in\mathbb{R}^{n}$ represents the filtered output. 
This formulation is often identified as the \textit{graph convolution}~\cite{DiscreteSignalProcessingonGraphs} operation. 
Due to the intensive computation cost associated with eigendecomposition, the mapping $f$ is typically implemented via polynomial approximations in practice, resulting in the derivation of Eq.~\ref{eq:graph-filtering} as below:
\begin{align}
\label{eq:graph-filtering-polynomial-graphwise}
\boldsymbol{z} = \boldsymbol{U}diag\left(\sum_{d=0}^{D}\theta_{d}\mathbf{T}_{d}(\boldsymbol{\lambda})\right)\boldsymbol{U}^{T}\boldsymbol{x} = \sum_{d=0}^{D}\theta_{d}\mathbf{T}_{d}(\boldsymbol{L})\boldsymbol{x}\ .
\end{align}
$\mathbf{T}_{d}$ denotes the $d$-th term of a polynomial, with coefficient $\theta_{d}$.

\subsubsection*{\bf Spectral-based GNNs} Spectral-based GNNs emerge from the integration of graph filters with graph-structured data. 
By treating each column of the node feature matrix $\boldsymbol{X}$ as an individual graph signal, a $L$-layer spectral GNN is architected as multi-layer neural network that processes the hidden feature through filtering operations, as formulated below~\cite{surveyspectralgnn}:
\begin{equation}
\label{eq:spectralgnns}\boldsymbol{H}^{(l+1)} = \sigma^{(l)}\left[\sum_{d=0}^{D}\theta_{dl}\mathbf{T}_{d}(\boldsymbol{L})\boldsymbol{H}^{(l)}\boldsymbol{W}^{(l)}\right],\quad \boldsymbol{H}^{(0)} \triangleq \boldsymbol{X}\ .
\end{equation}
Here, $\boldsymbol{H}^{(l)}$ and $\boldsymbol{W}^{(l)}$ correspond to the hidden layer representation and weight matrix at the $l$-th layer, respectively, with $\sigma^{(l)}$ representing a non-linear function commonly applied in neural networks. 
Each $l$-th layer is termed a \textit{graph convolution layer}, representing a critical building block in spectral GNNs and the subsequent developments in the field~\cite{ChebNet,ChebNetII,chebnet2d,decoupled-TrigoNet,ClenshawGCN,JacobiConv,OptBasisGNN,decoupled-AdaptKry,ERGNN,decoupled-PCConv,decoupled-UniFilter,decoupled-NFGNN}.


\section{Connecting Polynomial Capability with Spectral GNN Efficacy}
\label{section-connect-polynomial-ability-spectral-GNN-ability}

This section seeks to connect polynomial capability with the efficacy of spectral GNN. 
We examine the relationship between polynomial approximation errors and feature construction errors in graph convolution layer, providing theoretical analysis alongside intuitive numerical evaluations. 
This exploration yields vital insights that contribute to the progression of spectral GNNs in a polynomial context. 
We begin by defining several essential concepts.

\begin{definition}
\label{def:function-slices}
(\textit{Function slices}). Let $f: \left[0,2\right] \mapsto \mathbb{R}$ be a continuous and differentiable filter mapping. 
Denote the eigenvalues $\lambda_{1},\lambda_{2},...,\lambda_{n}$ of $\boldsymbol{L}$, satisfying $0=\lambda_{1}\leq\lambda_{2}\leq...\leq\lambda_{n}\leq2$. 
The function slices of $f(x)$ are given by a set of disjoint functions $f_{s}$, $s=1,2,...,n$, satisfying the following conditions:
\begin{equation}
\label{eq:function-slices-1}
f_{s}(x)=\left\{
\begin{array}{lc}
   f(x)  & x\in\left[\lambda_{s-1}, \lambda_{s}\right]\ , \\
    0  & Otherwise\ .
\end{array}
\right.
\end{equation}
Therefore, for any arbitrary function $f$, we can represent it by summing its slices, as illustrated below:
\begin{equation}
\label{eq:function-slices-2}
f(x)=\sum_{s=1}^{n}f_{s}(x)\ .
\end{equation}
\end{definition}
Figure~\ref{def:function-slices} provides an intuitive example of function slicing. 
This concept parallels the signal decomposition techniques found in the conventional signal processing field~\cite{Digitalsignalprocessing}.

\begin{figure}[!t]
\vskip -0.05in
\centering
\includegraphics[width=\linewidth]{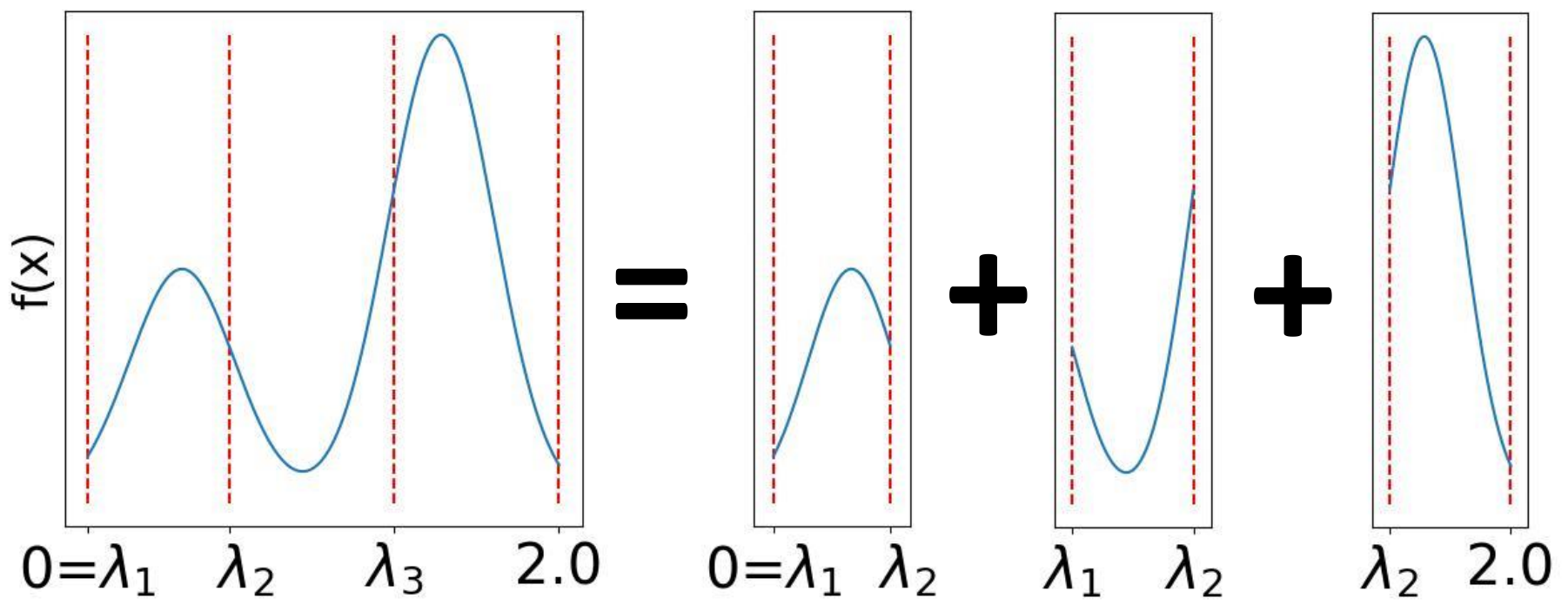}
\vskip -0.05in
\caption{Example of function slicing. 
$f(x)$ is dissected into three components, determined by its eigenvalues.}
\label{fig:slices}
\vskip -0.05in
\end{figure}

\begin{definition}
\label{def:polynomial-approximation-error}
(\textit{Polynomial's approximation error}). Let $\mathbf{T}_{0:D}(x;f)$ represent a polynomial function of degree $D$ that achieves the least squares error (LSE)~\cite{polyapprox_1,polyapprox_2} in approximating a specified function $f(x)$. 
Accordingly, we can define both continuous and discrete forms of the approximation error relative to the target filter function $f(x)$ using $\mathbf{T}_{0:D}$ as follows:
\begin{align}
\label{eq:polynomial-approximation-error-continuous}
\text{(Continuous)}\quad\quad &\epsilon \triangleq \int_{0}^{2}\lvert\mathbf{T}_{0:D}(x;f)-f(x)\rvert^{2}\ dx\ ,\\
\label{eq:polynomial-approximation-error-discrete}
\text{(Discrete)}\quad\quad &\epsilon \triangleq \lVert\mathbf{T}_{0:D}(\boldsymbol{\lambda};f)-f(\boldsymbol{\lambda})\rVert_{F}\ ,
\end{align}
where $\lVert\cdot\rVert_{F}$ denotes Frobenius norm~\cite{linearalgebra}.
\end{definition}
Our analysis centers on the continuous form, with derived insights adapted to the discrete form for application in spectral GNNs.

\begin{definition}
\label{def:graph-convolution-layer-construction-error}
(\textit{Construction error of graph convolution layer}). Let $\boldsymbol{Y}$ denote the target output of a graph convolution layer, expressed as $\boldsymbol{Y}=\boldsymbol{U}diag(f(\boldsymbol{\lambda}))\boldsymbol{U}^{T}\boldsymbol{X}\boldsymbol{W}$, where $f$ serves as the ``optimal'' filter function for constructing $\boldsymbol{Y}$. 
The construction error of the graph convolution layer on $\boldsymbol{Y}$ through a $D$-degree polynomial filter function $\mathbf{T}_{0:D}$ is defined as:
\begin{equation}
\label{eq:graph-convolution-layer-construction-error}
\xi \triangleq \lVert\boldsymbol{U}diag(\mathbf{T}_{0:D}(\boldsymbol{\lambda};f)-f(\boldsymbol{\lambda}))\boldsymbol{U}^{T}\boldsymbol{X}\boldsymbol{W}\rVert_{F}\ .
\end{equation}
\end{definition}
Note that the error $\xi$, analogous to $\epsilon$ in Definition~\ref{def:polynomial-approximation-error}, is measured as the difference between the target function $f$ and the polynomial $\mathbf{T}_{0:D}$ that achieves the least squares error (LSE) approximation. 
The graph convolution layer introduced in Definition~\ref{def:graph-convolution-layer-construction-error} aligns with a \textit{one-layer linear GNN}, a configuration similarly explored in previous studies~\cite{gnn-task-optimization,JacobiConv}. 
These prior works have examined the effectiveness of a one-layer linear GNN in constructing node labels to evaluate the overall performance of GNNs, which inspired us to examine the construction error within the graph convolution layer.

\subsection{Theoretical insights}
\label{section-connect-polynomial-ability-spectral-GNN-ability-theory}

Polynomial capability is quantified by the function approximation error~\cite{polyapprox_1,polyapprox_2}, whereas spectral GNN efficacy is typically reflected by prediction error in downstream tasks~\cite{GCN,GPRGNN,JacobiConv,decoupled-NFGNN,decoupled-FEGNN,decoupled-AdaptKry,decoupled-PCConv}. 
Consequently, a natural step toward linking polynomial capabilities with spectral GNN efficacy is to establish a bridge between the polynomial approximation error, $\epsilon$, as defined in Definition~\ref{def:polynomial-approximation-error}, and the graph convolution layer’s construction error, $\xi$.

In particular, as described in Definition~\ref{def:function-slices}, for an ``optimal'' filter function $f(x)$, the approximation error of a $D$-degree polynomial $\mathbf{T}_{0:D}(x;f)$ satisfies the conditions outlined in the following Lemma:
\begin{lemma}
\label{lemma:inequality-polynomial-error}
Let $f(x)$ be a function composed of function slices $f_{s}(x)$, $s=1,2,...,n$. 
Let $\mathbf{T}_{0:D}(x;f)$ be a $D$-degree polynomial that provides LSE approximation of $f(x)$ with error $\epsilon$. 
Specially, define $\epsilon_{s}$, $s=1,2,...,n$, as the polynomial approximation error of each slice $f_{s}(x)$ when approximated by the $D$-degree polynomial $\mathbf{T}_{0:D}(x;f_{s})$. 
An inequality that bounds $\epsilon$ in terms of $\epsilon_{s}$ are formulated below:
\begin{equation}
\label{eq:inequality-polynomial-error}
\sum_{s=1}^{n}\epsilon_{s} \leq \epsilon \leq (\sum_{s=1}^{n}\sqrt{\epsilon_{s}})^{2}\ .
\end{equation}
\end{lemma}
Proof can be found in Appendix. 
Lemma~\ref{lemma:inequality-polynomial-error} establishes both upper and lower bounds for the approximation error of a polynomial in relation to an arbitrary function $f$, based on the errors associated with its slices $f_{s}$. 
This result suggests that the capacity of the polynomial can be equivalently evaluated through the approximation error of its slices.

Drawing from the insights of bounded error above, we can now propose an inequality that bounds the construction error of the graph convolution layer, utilizing the polynomial approximation error as outlined in the theorem below:
\begin{theorem}
\label{theorem:inequality-gnn-layer-error}
Let $\delta_{\boldsymbol{X}}$ and $\delta_{\boldsymbol{W}}$ denote the minimum singular values of $\boldsymbol{X}$ and $\boldsymbol{W}$, respectively. 
Consider a regularization on the weight matrix $\boldsymbol{W}$, namely L$_2$ regularization, expressed as $\lVert\boldsymbol{W}\rVert_{F} \leq r$. 
The construction error $\xi$, satisfies the following inequality:
\begin{equation}
\label{eq:inequality-gnn-layer-error}
\delta_{\boldsymbol{X}}\delta_{\boldsymbol{W}}\sum_{s=1}^{n}\epsilon_{s} \leq \xi \leq r\lVert\boldsymbol{X}\rVert_{F}(\sum_{s=1}^{n}\sqrt{\epsilon_{s}})^{2}\ .
\end{equation}
\end{theorem}
Proof can be found in Appendix. 
Theorem~\ref{theorem:inequality-gnn-layer-error} establishes a direct connection between the polynomial approximation error and the construction error of the graph convolution layer through the approximation error of function slices, $\epsilon_{s}$. 
This insight is novel and, to our knowledge, has not been documented before.

\subsection{Numerical validation}
\label{section-connect-polynomial-ability-spectral-GNN-ability-exp}

\begin{figure}[!t]
\vskip -0.05in
\centering
    \subfloat[$f_{1}(x)$.]{
    \label{fig:f1}
    \includegraphics[width=0.3\linewidth]{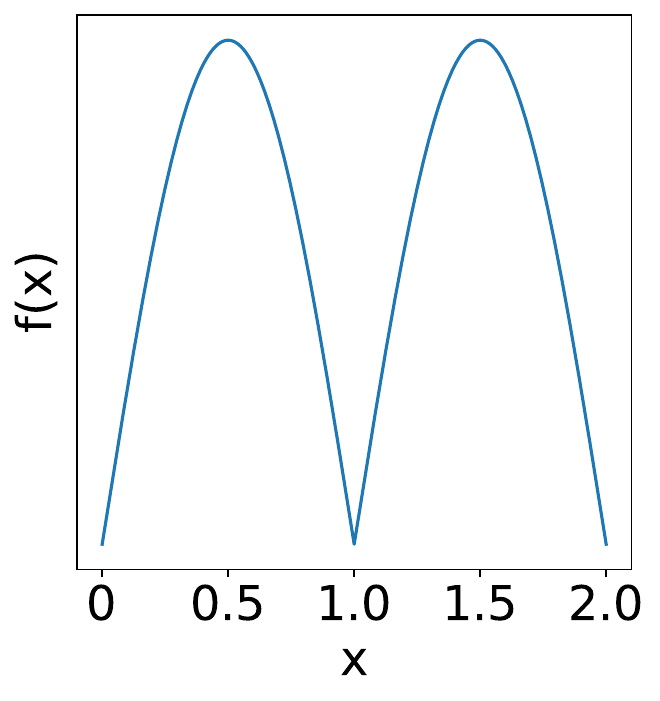}
    }
    \subfloat[$f_{2}(x)$.]{
    \label{fig:f2}
    \includegraphics[width=0.3\linewidth]{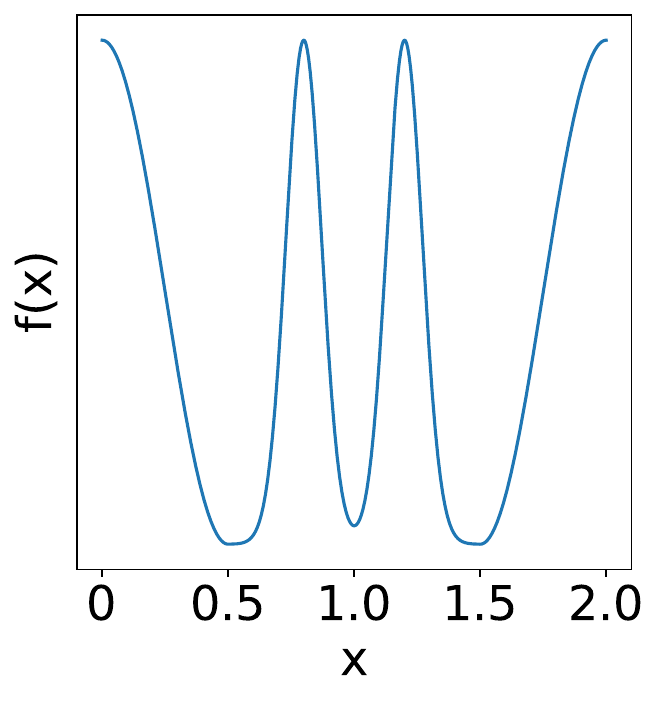}
    }
    \subfloat[$f_{3}(x)$.]{
    \label{fig:f3}
    \includegraphics[width=0.3\linewidth]{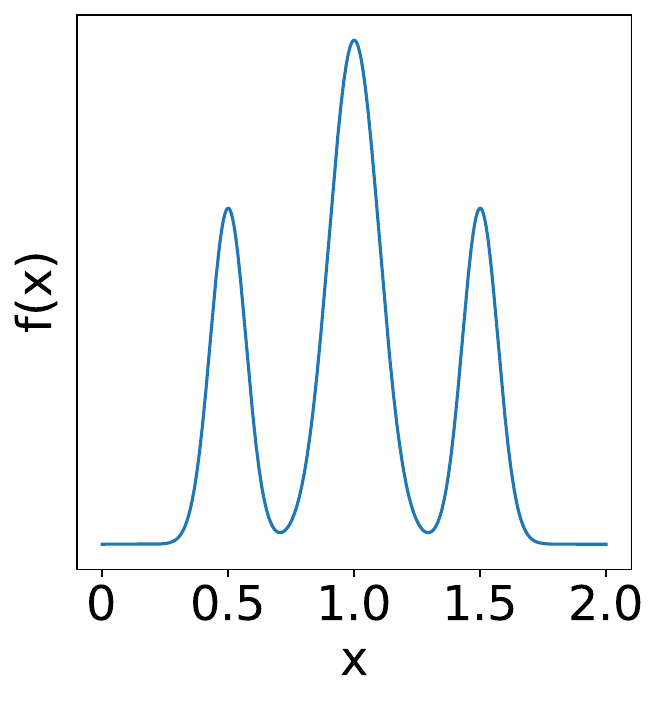}
    }\\ \vskip -0.05in
    \subfloat[$f_{4}(x)$.]{
    \label{fig:f4}
    \includegraphics[width=0.3\linewidth]{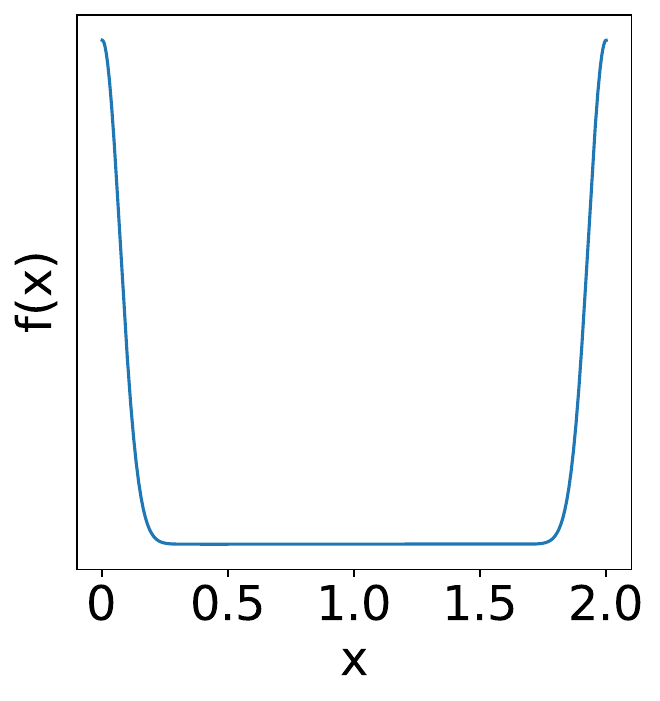}
    }
    \subfloat[$f_{5}(x)$.]{
    \label{fig:f5}
    \includegraphics[width=0.3\linewidth]{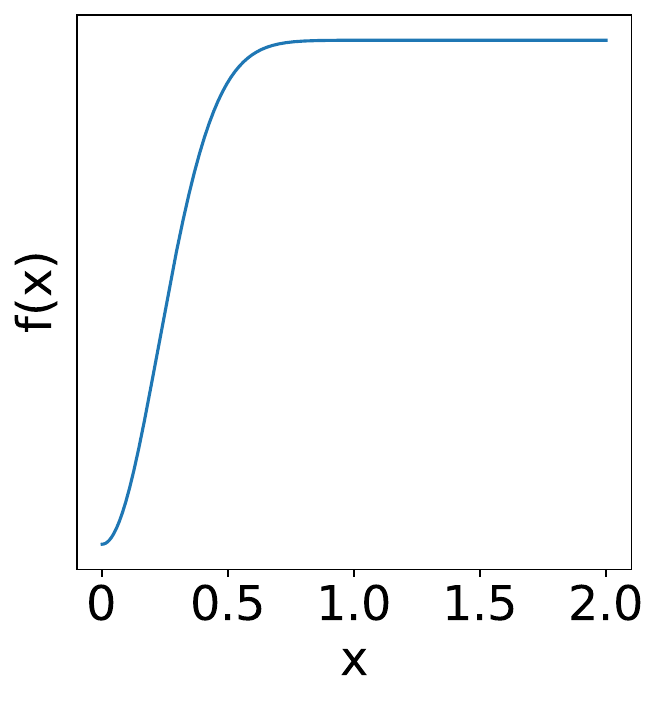}
    }
    \subfloat[$f_{6}(x)$.]{
    \label{fig:f6}
    \includegraphics[width=0.3\linewidth]{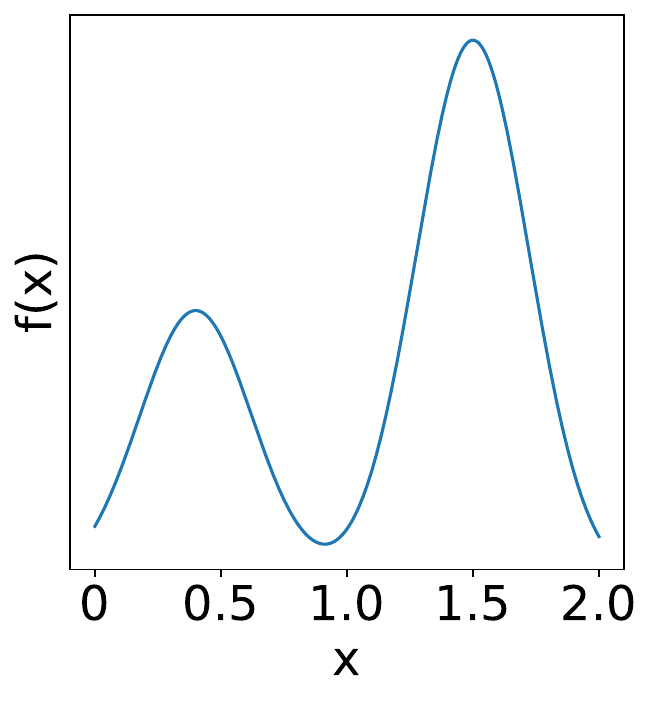}
    }
    \vskip -0.05in
    \caption{The functions served as target filters. 
    Additional mathematical details are available in Appendix.}
    \label{fig:filters}
\vskip -0.05in
\end{figure}

We conduct extensive numerical experiments to validate our theoretical findings. 
Inspired by filter learning experiments from prior spectral GNN studies~\cite{BernNet-GNN-narrowbandresults-1,decoupled-longnn,OptBasisGNN,decoupled-EC}, we design more challenging tasks with (i) increased graph sizes and (ii) complex target functions for learning. 
Specifically, we generate random graphs with $50,000$ nodes, substantially larger than the typical $10,000$-node setups in previous studies. 
Additionally, we utilize six intricate target filters, visualized in Figure~\ref{fig:filters}. 
The experiments comprise two primary tasks:
\begin{itemize}[leftmargin=*,parsep=0pt,itemsep=2pt,topsep=2pt]
\item Using eigenvalue-based slices of each function, we assess the approximation quality of five polynomials commonly adopted in spectral GNN literature, with the sum of squared errors (SSE) across $50000$ slices as the metric.
\item With a random $50000\times100$ matrix as node feature $\boldsymbol{X}$, we apply six target functions as filters, obtaining output $\boldsymbol{Y}_{1}$ to $\boldsymbol{Y}_{6}$. 
We train spectral GNNs on $(\boldsymbol{X},\boldsymbol{Y})$ to learn the target functions with polynomial filters, with the Frobenius norm of the difference between the learned and target filters as the metric.
\end{itemize}
\begin{table*}[!th]
\caption{Numerical experiment results.
\# Avg Rank 1 denotes the average rank in polynomial approximation, and \# Avg Rank 2 refers to the average rank in filter learning.} 
\vskip -0.05in
\label{table-numerical}
\centering
\setlength{\tabcolsep}{4pt}
\resizebox{\textwidth}{!}{
\begin{tabular}{cc|cccccc|cccccc|cc}
\hline
\multicolumn{2}{c|}{Method}                         & \multicolumn{6}{c|}{Slice-wise approximation}                               & \multicolumn{6}{c|}{Filter Learning}                                        & \multirow{2}{*}{\makecell[c]{\# Avg \\ Rank 1}} & \multirow{2}{*}{\makecell[c]{\# Avg \\ Rank 2}} \\ \cline{1-14}
Polynomial    & \multicolumn{1}{c|}{GNN} & $f_{1}(x)$ & $f_{2}(x)$ & $f_{3}(x)$ & $f_{4}(x)$ & $f_{5}(x)$ & $f_{6}(x)$ & $f_{1}(x)$ & $f_{2}(x)$ & $f_{3}(x)$ & $f_{4}(x)$ & $f_{5}(x)$ & $f_{6}(x)$ &                              &                              \\ \hline
Monomial      & GPRGNN~\cite{GPRGNN}                              & 139.9      & 289.1      & 466.1      & 398.3      & 1.83       & 97.83      & 167.2      & 366.4      & 566.3      & 468.7      & 15.91      & 139.2      & 5                            & 5                            \\
Bernstein     & BernNet~\cite{BernNet-GNN-narrowbandresults-1}                             & 32.78      & 247.3      & 398.5      & 306.5      & 0.058      & 22.92      & 68.23      & 313.2      & 448.2      & 415.2      & 7.79       & 95.84      & 4                            & 4                            \\
Chebyshev     & ChebNetII~\cite{ChebNetII}                           & 23.45      & 85.19      & 244.8      & 187.2      & 0.018      & 13.13      & 64.22      & 168.4      & 402.5      & 347.5      & 6.83       & 86.25      & 3                            & 3                            \\
Jacobian      & JacobiConv~\cite{JacobiConv}                          & 22.18      & 80.77      & 239.2      & 155.3      & 0.017      & 11.82      & 48.56      & 95.92      & 338.1      & 266.4      & 5.33       & 65.13      & 2                            & 2                            \\
Learnable     & OptBasis~\cite{OptBasisGNN}                            & 20.75      & 80.53      & 225.7      & 152.7      & 0.017      & 11.20      & 43.44      & 89.48      & 289.5      & 238.1      & 4.98       & 61.70      & 1                            & 1                            \\ \hline
\end{tabular}}
\vskip -0.05in
\end{table*}

\subsubsection*{\bf Numerical insights} Table~\ref{table-numerical} reveals that reducing the sum of the polynomial approximation error over function slices yields lower filter learning errors in spectral GNNs, consistently ranking both tasks. 
Although these results are derived from numerical experiments and may introduce certain biases, they confirm our theoretical analysis, showing a strong positive relationship between the polynomial’s capability and the efficacy of spectral GNNs.

\subsection{Summary}
\label{section-connect-polynomial-ability-spectral-GNN-ability-summary}
In this section, we summarize the significant findings from the preceding analysis and delve into discussions on enhancing spectral GNNs through the introduction of informed polynomial selection.

Specifically, as discussed in Section~\ref{section-connect-polynomial-ability-spectral-GNN-ability-theory} and~\ref{section-connect-polynomial-ability-spectral-GNN-ability-exp}, the construction error of spectral GNNs is intricately connected to the polynomial approximation error summed over function slices. 
Moreover, referring to Theorem~\ref{theorem:inequality-gnn-layer-error}, note that $\boldsymbol{X}$ is typically a constant property of the graph data, the construction error of graph convolution layer, $\xi$, therefore depends entirely on the slice-wise errors $\epsilon_{s}$, $s=1,2,...,n$. 
Consequently, for a graph data $\mathcal{G}=(\boldsymbol{A}, \boldsymbol{X})$ with node label $\boldsymbol{Y}$, an intuitive solution to reduce spectral GNN construction error is to utilize polynomials adept at approximating these slices.

Furthermore, practical graphs often contain millions of nodes~\cite{dataset5-ogb,dataset6-large-hetero} and complex target filters~\cite{GPRGNN,BernNet-GNN-narrowbandresults-1,splineGNN,decoupled-NewtonNet}. 
These characteristics result in very narrow and sharp slices of the target functions. 
Consequently, to minimize construction errors in spectral GNNs and improve their effectiveness, it is vital to incorporate ``narrow function-preferred'' polynomials in the development of graph filters. This insight not only represents a key contribution of this paper but also illuminates potential avenues for advancing spectral GNNs, paving the way for the introduction of a more advanced method.


\section{The proposed TFGNN}
\label{section-method}

Based on the previous analysis, this section introduces a novel trigonometric polynomial-based graph filter to enhance spectral GNNs. 
We begin with the trigonometric filter, discuss its efficient implementation via Taylor-based parameter decomposition, and present our {\bf T}rigonometric {\bf F}ilter {\bf G}raph {\bf N}eural {\bf N}etwork (TFGNN) as a decoupled GNN. 
A complexity analysis concludes the section.

\subsection{Parameter-efficient trigonometric filter}
\label{section-method-filter}

Trigonometric polynomials, among the most extensively utilized, have found widespread applications in approximating the functions with complicated patterns~\cite{Digitalsignalprocessing,TrigonometricSeries,trigo_interpolation_1}. 
More importantly, extensive prior studies in traditional signal processing domain have consistently highlighted the effectiveness of trigonometric polynomials over other polynomial types in modeling the functions localized within narrow intervals~\cite{Trigoapprox-narrowband-1,Trigoapprox-narrowband-2,Trigoapprox-narrowband-3,Trigoapprox-narrowband-4,Trigoapprox-narrowband-5,Trigoapprox-narrowband-6}. 
This prompts us to pioneer the development of graph filters through leveraging the power of trigonometric polynomials. 
Explicitly, the definition of our trigonometric graph filter is as follows: 
\begin{equation}
\label{eq:trigonometric-filter}
f_{\text{Trigo}}(\boldsymbol{\lambda})=\sum_{k=0}^{K}\left[\alpha_{k}\sin(k\omega\boldsymbol{\lambda})+\beta_{k}\cos(k\omega\boldsymbol{\lambda})\right] \ .
\end{equation}
Here, $K$ denotes the order of the truncated trigonometric polynomial. 
The coefficients $\alpha_{k}$, and $\beta_{k}$ are parameterized, while the hyperparameter $\omega$ (base frequency) is chosen from within the range $\left(0,\pi\right)$, enabling the trigonometric polynomial approximation to cover the interval $\left[0,2\right]$, which corresponds to the range of $\boldsymbol{\lambda}$. 
similar to  other types of polynomials, trigonometric polynomials offer considerable approximation capability for arbitrary functions, thus ensuring comprehensive filter coverage in practical applications~\cite{TrigonometricSeries,trigo_interpolation_1}.

\subsubsection*{\bf Guaranteed parameter-efficiency.} Apart from their recognized approximation capacities, trigonometric polynomials grant the graph filter $f_{\text{Trigo}}$ with a unique, provable efficiency regarding its parameters $(\alpha_{k},\beta_{k})$, $k\in\mathbb{N}$, as demonstrated in the theorem below.
\begin{theorem}
\label{theorem:finiteterms}
\textit{(Parameter-efficiency)}. Given a $f(x)$ formulated as $f_{\text{Trigo}}(x)$, its coefficients $\alpha_{k}$ and $\beta_{k}$ satisfies:
\begin{equation}
\label{eq:finiteterms}
\lim_{k\to +\infty}\alpha_{k}=0 \ , \quad \  \lim_{k\to +\infty}\beta_{k}=0 \ .
\end{equation}
\end{theorem}
Proof can be found in Appendix. 
Theorem~\ref{theorem:finiteterms} establishes a solid basis by revealing that polynomial terms with larger values of $k$ within $f_{\text{Trigo}}(x)$ correspond to smaller weights. 
This insight indicates that the contribution of high-order terms is relatively insignificant, allowing for their practical omission without substantial loss in approximation accuracy. 
As a result, $f_{\text{Trigo}}(x)$ can achieve substantial effectiveness with only a small $K$, reducing the complexity of the filters while retaining significant accuracy. 
This reinforces the superiority of $f_{\text{Trigo}}(x)$ over other graph filter designs.

\subsection{Taylor-based parameter decomposition}
\label{section-method-parameter-decomposition}

As detailed in Eq.~\ref{eq:trigonometric-filter}, implementing the standard $f_{\text{Trigo}}(x)$ requires an eigen-decomposition, which imposes substantial computational complexity and limits scalability compared to alternative methods. 
We tackle this challenge through the introduction of Taylor-based parameter decomposition (TPD). 
TPD first reformulates the trigonometric terms $\sin(k\omega\boldsymbol{\lambda})$ and $\cos(k\omega\boldsymbol{\lambda})$ into polynomial forms through the Taylor expansion~\cite{numericalexp,polyapprox_1,polyapprox_2}, as shown below:
\begin{equation}
\sin(k\omega\boldsymbol{\lambda}) = \sum_{d=0}^{D}\gamma_{kd}\boldsymbol{\lambda}^{d},\ \cos(k\omega\boldsymbol{\lambda}) = \sum_{d=0}^{D}\theta_{kd}\boldsymbol{\lambda}^{d}\ .
\end{equation}
The constants $\gamma_{kd}$ and $\theta_{kd}$ depend exclusively on the types of functions (sine and cosine), the index $k$, and the hyperparameter $\omega$. 
The effectiveness of Taylor expansion for modeling functions within localized intervals has been thoroughly established in the literature~\cite{taylor_app_1,taylor_app_2,taylor_app_3,taylor_app_4,taylor_app_5,taylor_app_6}, especially for trigonometric functions~\cite{taylor_trigo_1,taylor_trigo_2,taylor_trigo_3}. 
Since $\boldsymbol{\lambda}$ is restricted to the range $\left[0,2\right]$, the Taylor expansion emerges as a viable and crucial strategy for efficiently approximating these trigonometric functions.

With the updated formulations, TPD alters the convolution operation with $f_{\text{Trigo}}(x)$ on the node feature $\boldsymbol{X}$ as detailed below:
\begin{align}
\boldsymbol{Z} 
&= \boldsymbol{U}\sum_{k=0}^{K}\left[\alpha_{k}\sum_{d=0}^{D}\gamma_{kd}diag(\boldsymbol{\lambda}^{d})+\beta_{k}\sum_{d=0}^{D}\theta_{kd}diag(\boldsymbol{\lambda}^{d})\right]\boldsymbol{U}^{T}\boldsymbol{X}\ , \notag \\
\label{eq:graph-conv-TPD}
&= \sum_{d=0}^{D}\boldsymbol{L}^{d}\boldsymbol{X}\left(\boldsymbol{\alpha}\boldsymbol{\Gamma}_{:d} + \boldsymbol{\beta}\boldsymbol{\Theta}_{:d}\right)\ .
\end{align}
Here, $\boldsymbol{\alpha}$ and $\boldsymbol{\beta}$ denote the $K+1$-dimensional vectors with elements being $\alpha_{k}$ and $\beta_{k}$, respectively. 
$\boldsymbol{\Gamma}$ and $\boldsymbol{\Theta}$ refer to the $(K+1)\times(D+1)$ matrices formed with $\gamma_{kd}$, $\theta_{kd}$. 
Eq.~\ref{eq:graph-conv-TPD} illustrates a streamlined convolution with $f_{\text{Trigo}}(x)$, offering two significant benefits:
\begin{itemize}[leftmargin=*,parsep=0pt,itemsep=2pt,topsep=2pt]
\item {\bf Reduced complexity.} Utilizing the TPD, the graph convolution with $f_{\text{Trigo}}(x)$ eliminates the need for computation-intensive eigen-decomposition. 
This reduction in computational overhead brings the costs in line with those of standard polynomial-based filters, leading to significant efficiency gains.
\item {\bf Parameter decomposition.} TPD integrates all trigonometric functions into polynomial forms, allowing for increases in $K$ to only influence trivial computations of $\boldsymbol{\alpha}\boldsymbol{\Gamma}$ and $\boldsymbol{\beta}\boldsymbol{\Theta}$, enhancing precision of $f_{\text{Trigo}}(x)$ with negligible additional cost.
\end{itemize}
%

\subsection{Modeling TFGNN as decoupled paradigm}
\label{section-method-decoupled}

TFGNN is a decoupled GNN architecture that separates graph convolution from feature transformation. 
This design principle, first proposed by~\cite{decoupled-advantages-3-coupled-disadvantages-2-APPNP}, has become a de facto choice in modern spectral GNNs for its significant efficacy and computational efficiency~\cite{GPRGNN,BernNet-GNN-narrowbandresults-1,EvenNet,chebnet2d,ChebNetII,OptBasisGNN,JacobiConv,decoupled-UniFilter,decoupled-PCConv,decoupled-AdaptKry}, and even stands out as a promising solution for scalable GNNs~\cite{decoupled-LD2,decoupled-SCARA}. 
Specifically, incorporating the trigonometric filter $f_{\text{Trigo}}(x)$ with the introduced Taylor-based parameter decomposition, we present two versions of TFGNN to cater to different graph sizes:

\ding{182}~In the case of \textbf{medium-to-large} graphs like Cora~\cite{dataset1-cora} and Arxiv~\cite{dataset5-ogb}, TFGNN operates as described below:
\begin{align}
\label{eq:decoupled-TFGNN-medium}
\boldsymbol{Z}=\sum_{d=0}^{D}\boldsymbol{L}^{d}\boldsymbol{H}\left(\boldsymbol{\alpha}\boldsymbol{\Gamma}_{:d} + \boldsymbol{\beta}\boldsymbol{\Theta}_{:d}\right),\quad \boldsymbol{H}=\mathrm{MLP}(\boldsymbol{X})\ .
\end{align}
$\mathrm{MLP}(\cdot)$ denotes a multi-layer perceptron for feature transformation. 

\ding{183}~In the case of \textbf{exceptionally large} graphs, such as Wiki~\cite{dataset6-large-hetero} and Papers100M~\cite{dataset5-ogb}, TFGNN is implemented as follows:
\begin{align}
\label{eq:decoupled-TFGNN-large}
\boldsymbol{Z}=\mathrm{MLP}(\boldsymbol{H}),\quad \boldsymbol{H}=\sum_{d=0}^{D}\boldsymbol{L}^{d}\boldsymbol{X}\left(\boldsymbol{\alpha}\boldsymbol{\Gamma}_{:d} + \boldsymbol{\beta}\boldsymbol{\Theta}_{:d}\right)\ ,
\end{align}
The different implementations of TFGNN come from hardware constraints and introduce notable benefits: 
(i) for medium-to-large graphs, the graph data can be fully stored on GPUs; 
therefore, by simply reducing the feature dimensions with MLP, the subsequent convolution process could achieves high efficiency; 
(ii) for exceptionally large graphs, where GPU memory limitations become a substantial challenge, TFGNN precomputes features, $\boldsymbol{L}^{d}\boldsymbol{X}$, $d=1,2,...D$, and stores them as static data files. 
This allows for efficient graph convolution operation via repeated reads of precomputed features, mitigating the intense computational complexity associated with GNN training; an efficient MLP is applied later.

\begin{table}[!t]
\caption{Complexity comparison of TFGNN against others. 
The complexity pertains to the graph convolution layers.}
\vskip -0.05in
\label{table-complexity}
\centering
\begin{tabular}{c|c|c}
\hline
Method & Computation & Parameter \\ \hline
GPRGNN~\cite{GPRGNN} & $\mathcal{O}(mED)$ & $\mathcal{O}(K)$ \\
ChebNetII~\cite{ChebNetII} & $\mathcal{O}(mED)$ & $\mathcal{O}(K)$ \\ 
OptBasis~\cite{OptBasisGNN} & $\mathcal{O}(mED)$ & $\mathcal{O}(K)$ \\ 
JacobiConv~\cite{JacobiConv} & $\mathcal{O}(mED)$ & $\mathcal{O}(K)$ \\ 
UniFilter~\cite{decoupled-UniFilter} & $\mathcal{O}(mED)$ & $\mathcal{O}(K)$ \\ 
TFGNN ({\bf ours}) & $\mathcal{O}(mED)$ & $\mathcal{O}(K)$  \\ \hline
\end{tabular}
\vskip -0.05in
\end{table}

\subsection{Complexity analysis of TFGNN}
\label{section-TFGNN-complexity}

This subsection presents the complexity analysis of TFGNN, with a particular emphasis on graph convolution layers, as the complexity of feature transformation MLPs is already well-understood.
To start, we consider a graph $\mathcal{G}$ with $n$ nodes, $E$ edges, and $m$ feature dimensions. 
Across all spectral GNNs, the maximum polynomial order is set to $D$. 
The trigonometric polynomial degree is capped at $K$. 
A summary of the complexity analysis is outlined in Table~\ref{table-complexity}.

\subsubsection*{\bf Computational complexity.} As shown in Eq.~\ref{eq:graph-conv-TPD}, for each order $d$, TFGNN first computes $\boldsymbol{\alpha}\boldsymbol{\Gamma}_{:d}$ and $\boldsymbol{\beta}\boldsymbol{\Theta}_{:d}$, requiring $2(K+1)$ operations, followed by propagating with $\boldsymbol{L}$, which requires $mE$ computations. 
Since the number of edges $E$ is millions of times larger than both $K$ and $D$, the practical complexity is governed by $mE$ for each order $d$, leading to an overall complexity of $\mathcal{O}(mED)$. 
This is comparable to other spectral GNNs like GPRGNN~\cite{GPRGNN}, which utilizes recursive computation of the propagated feature $\boldsymbol{L}^{d}\boldsymbol{X}$. 
Thus, our TFGNN achieves complexity on par with prior methods. 
Additionally, for exceptionally large graphs, TFGNN reduces complexity further by precomputing all $\boldsymbol{L}^{d}\boldsymbol{X}$, thus avoiding redundant repeated computations during training.

\subsubsection*{\bf Parameter complexity.} Our TFGNN achieves a parameter complexity of $\mathcal{O}(2(K+1))$, in contrast to traditional spectral GNNs’ $\mathcal{O}(K+1)$, where each polynomial basis order is assigned a parameterized coefficient. 
This increase is, however, trivial, as the feature transformation MLP constitutes the majority of parameters, greatly outweighing the graph convolution layers. 
As such, TFGNN’s parameter complexity remains effectively on par with that of other spectral GNNs when considering the entire model.

\begin{table*}[!th]
\caption{Node classification results on medium-to-large graphs. 
\#Improv. denotes the performance gain of TFGNN over the best baseline result. 
Boldface represents the first result, while underlined indicates the runner-up.}
\vskip -0.1in
\label{table-node-classify-medium}
\centering
\setlength{\tabcolsep}{4.5pt}
\resizebox{\linewidth}{!}{
\begin{tabular}{cccccc|ccccc}
\hline
\multirow{2}{*}{\makecell[c]{GNN \\ Type}} & \multirow{2}{*}{Method}              & \multicolumn{4}{c|}{homophilic graphs}                                                                                        & \multicolumn{5}{c}{heterophilic graphs}                                                                                                                       \\ \cline{3-11} 
                      &                                      & Cora                          & Cite.                         & Pubmed                        & Arxiv                         & Roman.                        & Amazon.                       & Ques.                         & Gamers                        & Genius                        \\ \hline
\multirow{5}{*}{\large \romannumeral1}    & H2GCN                                & $87.33_{\pm0.6}$              & $75.11_{\pm1.2}$              & $88.39_{\pm0.6}$              & $71.93_{\pm0.4}$              & $61.38_{\pm1.2}$              & $37.17_{\pm0.5}$              & $64.42_{\pm1.3}$              & $64.71_{\pm0.4}$              & $90.12_{\pm0.2}$              \\
                      & GLOGNN                               & $88.12_{\pm0.4}$              & $76.23_{\pm1.4}$              & $88.83_{\pm0.2}$              & $72.08_{\pm0.3}$              & $71.17_{\pm1.2}$              & $42.19_{\pm0.6}$              & $74.42_{\pm1.3}$              & $65.62_{\pm0.3}$              & $90.39_{\pm0.3}$              \\
                      & LINKX                                & $84.51_{\pm0.6}$              & $73.25_{\pm1.5}$              & $86.36_{\pm0.6}$              & $71.14_{\pm0.2}$              & $67.55_{\pm1.2}$              & $41.57_{\pm0.6}$              & $63.85_{\pm0.8}$              & $65.82_{\pm0.4}$              & $\underline{91.12_{\pm0.5}}$  \\
                      & OrderGNN                             & $87.55_{\pm0.2}$              & $75.46_{\pm1.2}$              & $88.31_{\pm0.3}$              & $71.90_{\pm0.5}$              & $71.69_{\pm1.6}$              & $40.93_{\pm0.5}$              & $70.82_{\pm1.0}$              & $66.09_{\pm0.3}$              & $89.45_{\pm0.4}$              \\
                      & LRGNN                                & $87.48_{\pm0.3}$              & $75.29_{\pm1.0}$              & $88.65_{\pm0.4}$              & $71.69_{\pm0.3}$              & $72.35_{\pm1.4}$              & $\underline{42.56_{\pm0.4}}$  & $71.82_{\pm1.1}$              & $66.29_{\pm0.5}$              & $90.38_{\pm0.7}$              \\ \hline
\multirow{5}{*}{\large \romannumeral2}   & GCN                                  & $86.48_{\pm0.4}$              & $75.23_{\pm1.0}$              & $87.29_{\pm0.2}$              & $71.77_{\pm0.1}$              & $72.33_{\pm1.6}$              & $42.09_{\pm0.6}$              & $75.17_{\pm0.8}$              & $63.29_{\pm0.5}$              & $86.73_{\pm0.5}$              \\
                      & GCNII                                & $86.77_{\pm0.2}$              & $\underline{76.57_{\pm1.5}}$ & $88.86_{\pm0.4}$              & $71.72_{\pm0.4}$              & $71.62_{\pm1.7}$              & $40.89_{\pm0.4}$              & $72.32_{\pm1.0}$              & $65.11_{\pm0.3}$              & $90.60_{\pm0.6}$              \\
                      & ChebNet                              & $86.83_{\pm0.7}$              & $74.39_{\pm1.3}$              & $85.92_{\pm0.5}$              & $71.52_{\pm0.3}$              & $64.44_{\pm1.5}$              & $38.81_{\pm0.7}$              & $70.42_{\pm1.2}$              & $63.62_{\pm0.4}$              & $87.42_{\pm0.2}$              \\
                      & ACMGCN                               & $87.21_{\pm0.4}$              & $76.03_{\pm1.4}$              & $87.37_{\pm0.4}$              & $71.70_{\pm0.3}$              & $66.48_{\pm1.2}$              & $39.53_{\pm0.9}$              & $67.84_{\pm0.5}$              & $64.73_{\pm0.3}$              & $83.45_{\pm0.7}$              \\
                      & Specformer                           & $88.19_{\pm0.6}$              & $75.87_{\pm1.5}$              & $88.74_{\pm0.2}$              & $71.88_{\pm0.2}$              & $71.69_{\pm1.4}$              & $42.06_{\pm0.8}$              & $70.75_{\pm1.2}$              & $65.80_{\pm0.2}$              & $89.39_{\pm0.6}$              \\
                      \hline
\multirow{8}{*}{\large \romannumeral3}  
                      & GPRGNN                               & $88.26_{\pm0.5}$              & $76.24_{\pm1.2}$              & $88.81_{\pm0.2}$              & $71.89_{\pm0.2}$              & $64.49_{\pm1.6}$              & $41.48_{\pm0.6}$              & $64.58_{\pm1.2}$              & $66.23_{\pm0.1}$              & $90.92_{\pm0.6}$              \\
                      & BernNet                              & $87.57_{\pm0.4}$              & $75.81_{\pm1.8}$              & $88.48_{\pm0.3}$              & $71.72_{\pm0.3}$              & $65.44_{\pm1.4}$              & $40.74_{\pm0.7}$              & $65.53_{\pm1.6}$              & $65.74_{\pm0.3}$              & $89.75_{\pm0.3}$              \\
                      & ChebNetII                            & $88.17_{\pm0.4}$              & $76.41_{\pm1.3}$              & $88.98_{\pm0.4}$              & $72.13_{\pm0.3}$              & $66.77_{\pm1.2}$              & $42.44_{\pm0.9}$              & $71.28_{\pm0.6}$              & $66.44_{\pm0.5}$              & $90.60_{\pm0.2}$              \\
                      & OptBasis                             & $88.35_{\pm0.6}$              & $76.22_{\pm1.4}$              & $89.38_{\pm0.3}$              & $72.10_{\pm0.2}$              & $64.28_{\pm1.8}$              & $41.63_{\pm0.8}$              & $69.60_{\pm1.2}$              & $\underline{66.81_{\pm0.4}}$  & $90.97_{\pm0.5}$              \\
                      & JacobiConv                           & $\underline{88.53_{\pm0.8}}$  & $76.27_{\pm1.3}$              & $\underline{89.51_{\pm0.2}}$  & $71.87_{\pm0.3}$              & $70.10_{\pm1.7}$              & $42.18_{\pm0.4}$              & $72.16_{\pm1.3}$              & $64.17_{\pm0.3}$              & $89.32_{\pm0.5}$              \\
                      & NFGNN                                & $88.06_{\pm0.4}$              & $76.22_{\pm1.4}$              & $88.43_{\pm0.4}$              & $72.15_{\pm0.3}$              & $\underline{72.46_{\pm1.2}}$  & $42.19_{\pm0.3}$              & $\underline{75.49_{\pm0.9}}$  & $66.64_{\pm0.4}$              & $90.87_{\pm0.5}$              \\
                      & AdaptKry                             & $88.23_{\pm0.7}$              & $76.54_{\pm1.2}$  & $88.38_{\pm0.6}$              & $72.33_{\pm0.3}$              & $71.40_{\pm1.3}$              & $42.31_{\pm1.1}$              & $72.55_{\pm1.0}$              & $66.27_{\pm0.3}$              & $90.55_{\pm0.3}$              \\
                      & UniFilter                            & $88.31_{\pm0.7}$              & $76.38_{\pm1.1}$              & $89.30_{\pm0.4}$              & $\underline{72.87_{\pm0.4}}$  & $71.22_{\pm1.5}$              & $41.37_{\pm0.6}$              & $73.83_{\pm0.8}$              & $65.75_{\pm0.4}$              & $90.66_{\pm0.2}$              \\ \hline
\multirow{2}{*}{}     & TFGNN (\textbf{Ours}) & \textbf{89.21}$\boldsymbol{_{\pm0.4}}$ & \textbf{77.68}$\boldsymbol{_{\pm0.8}}$              & \textbf{90.00}$\boldsymbol{_{\pm0.2}}$ & \textbf{75.23}$\boldsymbol{_{\pm0.2}}$ & \textbf{74.94}$\boldsymbol{_{\pm1.1}}$ & \textbf{45.04}$\boldsymbol{_{\pm0.6}}$ & \textbf{81.55}$\boldsymbol{_{\pm0.9}}$ & \textbf{69.46}$\boldsymbol{_{\pm0.2}}$ & \textbf{92.40}$\boldsymbol{_{\pm0.2}}$ \\
                      & \#Improv.                            & $0.68\%$                      & $1.11\%$                     & $0.49\%$                      & $2.36\%$                      & $2.48\%$                      & $2.48\%$                      & $6.06\%$                      & $2.65\%$                      & $1.28\%$                      \\ \hline
\end{tabular}}
\vskip -0.05in
\end{table*}


\section{Empirical Studies}
\label{section-exp}

This section details the empirical evaluations, including numerical experiments same as those in Section~\ref{section-connect-polynomial-ability-spectral-GNN-ability-exp}, a benchmark node classification task, and a practical application in graph anomaly detection. 
A demo code implementation is available through the GitHub repository - \url{https://github.com/vasile-paskardlgm/TFGNN}.

\subsection{Slice approximation and filter learning}
\label{section-exp-numerical}

We conduct the same numerical experiments as outlined in Section~\ref{section-connect-polynomial-ability-spectral-GNN-ability-exp} to evaluate the proposed trigonometric graph filters and TFGNN. 
To ensure a fair and informative comparison, the trigonometric polynomial used in our numerical experiments is not in its naive form; rather, we employ the formulation that incorporates a $10$ degree Taylor-based parameter decomposition, akin in that of Section~\ref{section-method-parameter-decomposition}. 
The polynomial degree $K$ is set to $10$, yielding $K+1$ coefficients in total. 
Thus, TFGNN preserves both the maximum order of $\boldsymbol{\lambda}$ and the number of coefficients used by other counterparts.

\subsubsection*{\bf Results} Table~\ref{table-numerical-trigo} summarizes the performance of TFGNN alongside the three leading alternatives—Chebyshev, Jacobian, and Learnable—with boldface marking the highest scores due to space constraints. 
More results can be found in Appendix. 
According to these results, trigonometric polynomials and TFGNN consistently outperform other methods. 
Particularly, for target functions exhibiting complex patterns, such as $f_{2}(x)$, $f_{3}(x)$, and $f_{4}(x)$, TFGNN obtains notable improvements over competitors, showing the efficacy of our method.
The following sections will show how TFGNN attains leading performance on real-world datasets, affirming that the numerical outcomes correspond well to real-world scenarios.

\begin{table}[!t]
\caption{Results of slice approximation and filter learning.} 
\vskip -0.1in
\label{table-numerical-trigo}
\centering
\setlength{\tabcolsep}{3pt}
\resizebox{\linewidth}{!}{
\begin{tabular}{ccccc|ccc}
\hline
\multicolumn{2}{c}{Method} & \multicolumn{3}{c|}{Poly. approx.}   & \multicolumn{3}{c}{Filter Learn.}    \\ \hline
Poly.      & GNN           & $f_{2}(x)$ & $f_{3}(x)$ & $f_{4}(x)$ & $f_{2}(x)$ & $f_{3}(x)$ & $f_{4}(x)$ \\ \hline
Cheby.     & ChebNetII     & 85.19      & 244.8      & 187.2      & 168.4      & 402.5      & 347.5      \\
Jacobi.    & JacobiConv    & 80.77      & 239.2      & 155.3      & 95.92      & 338.1      & 266.4      \\
Learn.     & OptBasis      & 80.53      & 225.7      & 152.7      & 89.48      & 289.5      & 238.1      \\ \hline
\textbf{Trigo.}     & \textbf{TFGNN}         & \textbf{23.69}      & \textbf{71.13}      & \textbf{59.88}      & \textbf{65.19}      & \textbf{102.3}      & \textbf{105.3}      \\ \hline
\end{tabular}}
\vskip -0.1in
\end{table}


\begin{table*}[!th]
\caption{Node classification and runtime (\textit{hours}) results on exceptionally large graphs. 
``OOM'' denotes ``Out-Of-Memory''.}
\vskip -0.1in
\label{table-node-classify-large}
\centering
\begin{tabular}{ccccccccc}
\hline
\multirow{2}{*}{Method} & \multicolumn{2}{c}{Products}            & \multicolumn{2}{c}{Papers100M}          & \multicolumn{2}{c}{Snap}                & \multicolumn{2}{c}{Pokec}               \\ \cline{2-9} 
                        & Test acc                      & Runtime & Test acc                      & Runtime & Test acc                      & Runtime & Test acc                      & Runtime \\ \hline
GCN                     & $76.37_{\pm0.2}$              & $1.2$   & OOM                           & -       & $46.66_{\pm0.1}$              & $1.9$   & $74.78_{\pm0.2}$              & $1.2$   \\
SGC                     & $75.16_{\pm0.2}$              & $0.9$   & $64.02_{\pm0.2}$              & $10.2$  & $31.11_{\pm0.2}$              & $1.6$   & $60.29_{\pm0.1}$              & $0.9$   \\
GPRGNN                  & $79.45_{\pm0.1}$              & $1.3$   & $66.13_{\pm0.2}$              & $11.1$  & $48.88_{\pm0.2}$              & $2.0$   & $79.55_{\pm0.3}$              & $1.2$   \\
BernNet                 & $79.82_{\pm0.2}$              & $1.3$   & $66.08_{\pm0.2}$              & $11.2$  & $47.48_{\pm0.3}$              & $2.1$   & $80.55_{\pm0.2}$              & $1.3$   \\
ChebNetII               & $81.66_{\pm0.3}$              & $1.2$   & $\underline{67.11_{\pm0.2}}$  & $11.0$  & $51.74_{\pm0.2}$              & $1.9$   & $81.88_{\pm0.3}$              & $1.2$   \\
JacobiConv              & $79.35_{\pm0.2}$              & $1.0$   & $65.45_{\pm0.2}$              & $10.5$  & $50.66_{\pm0.2}$              & $1.7$   & $73.83_{\pm0.2}$              & $1.0$   \\
OptBasis                & $81.33_{\pm0.2}$              & $1.3$   & $67.03_{\pm0.3}$              & $11.2$  & $53.55_{\pm0.1}$              & $2.1$   & $82.09_{\pm0.3}$              & $1.3$   \\
NFGNN                   & $81.11_{\pm0.2}$              & $1.3$   & $66.38_{\pm0.2}$              & $11.3$  & $\underline{57.83_{\pm0.3}}$  & $2.1$   & $81.56_{\pm0.3}$              & $1.4$   \\ 
AdaptKry                & $\underline{81.70_{\pm0.3}}$  & $1.4$   & $67.07_{\pm0.2}$              & $11.3$  & $55.92_{\pm0.2}$              & $2.1$   & $82.16_{\pm0.2}$              & $1.4$   \\
UniFilter               & $80.33_{\pm0.2}$              & $1.2$   & $66.79_{\pm0.3}$              & $11.0$  & $52.06_{\pm0.1}$              & $2.1$   & $\underline{82.23_{\pm0.3}}$  & $1.3$   \\ 
\hline
TFGNN (\textbf{Ours})           & \textbf{84.05}$\boldsymbol{_{\pm0.2}}$ & $1.2$   & \textbf{68.65}$\boldsymbol{_{\pm0.2}}$ & $11.0$  & \textbf{64.38}$\boldsymbol{_{\pm0.2}}$ & $1.9$   & \textbf{85.55}$\boldsymbol{_{\pm0.2}}$ & $1.2$   \\
\#Improv.               & $2.35\%$   &   -        & $1.54\%$     &    -     & $6.55\%$    &    -      & $3.32\%$     &     -    \\ \hline
\end{tabular}
\vskip -0.05in
\end{table*}


\subsection{Benchmark node classification tasks}
\label{section-exp-node-classify}

We further assess TFGNN via benchmark node classification tasks. 

\subsubsection{\bf Datasets and baselines}\
\subsubsection*{\bf Datasets.} We utilize $13$ benchmark datasets with varied sizes and heterophily levels~\cite{heterophily-gnn-survey}. 
For homophilic datasets, we comprise citation graphs (Cora, CiteSeer, PubMed)\cite{dataset1-cora} and large OGB graphs (ogbn-Arxiv, ogbn-Products, ogbn-Papers100M)\cite{dataset5-ogb}. 
For heterophilic datasets, we select three latest datasets (Roman-empire, Amazon-ratings, Questions)~\cite{dataset8-small-hetero} and four large ones (Gamers, Genius, Snap-patent, Pokec)~\cite{dataset6-large-hetero}. 
(We exclude conventional dataset choices~\cite{dataset3-pei} due to the recognized data-leakage issues in these datasets~\cite{dataset8-small-hetero}.) 

\subsubsection*{\bf Baselines and settings} We include $18$ advanced baselines tailored for both heterophilic and homophilic scenarios, which can be categorized into three classes as follows:
\begin{itemize}[leftmargin=*,parsep=0pt,itemsep=2pt,topsep=2pt]
\item {\bf Non-spectral GNNs:} H2GCN~\cite{H2GCN}, GLOGNN~\cite{glognn++}, LINKX~\cite{dataset6-large-hetero}, OrderGNN~\cite{OrderedGNN}, LRGNN~\cite{low-rank-gnn-2}. 
\item {\bf Non-decoupled spectral GNNs:} GCN~\cite{GCN}, GCNII~\cite{gcnii}, ChebNet~\cite{ChebNet}, ACMGCN~\cite{ACMGCN}, Specformer~\cite{specformer}. 
\item {\bf Decoupled spectral GNNs:} GPRGNN~\cite{GPRGNN}, BernNet~\cite{BernNet-GNN-narrowbandresults-1}, ChebNetII~\cite{ChebNetII}, OptBasis~\cite{OptBasisGNN}, NFGNN~\cite{decoupled-NFGNN}, JacobiConv~\cite{JacobiConv}, AdaptKry~\cite{decoupled-AdaptKry}, UniFilter~\cite{decoupled-UniFilter}.
\end{itemize}
For the widely adopted baselines (GCN and ChebNet), we adopt consistent implementations drawn from previous research~\cite{BernNet-GNN-narrowbandresults-1,ChebNetII,chebnet2d,OptBasisGNN,JacobiConv,decoupled-FEGNN,decoupled-AdaptKry,decoupled-NFGNN}. 
For the remaining baselines, we inherit the hyperparameter tuning settings from their original publications.

\subsubsection*{\bf Implementation of TFGNN} To ensure experimental fairness, we fix the order of the Trigonometric Polynomial Decomposition, denoted as $D$, to $10$, aligning with other baselines such as GPRGNN and ChebNetII. 
We employ a grid search to optimize the parameters $\omega$ within $\{0.2\pi,0.3\pi,0.5\pi,0.7\pi\}$ and $K$ from $\{2,4,6,8,10,15,20\}$. 
Additional details are in Appendix.

\begin{figure}[!t]
\vskip -0.05in
\centering
    \subfloat[Cora.]{
    \label{fig:cora}
    \includegraphics[width=0.45\linewidth]{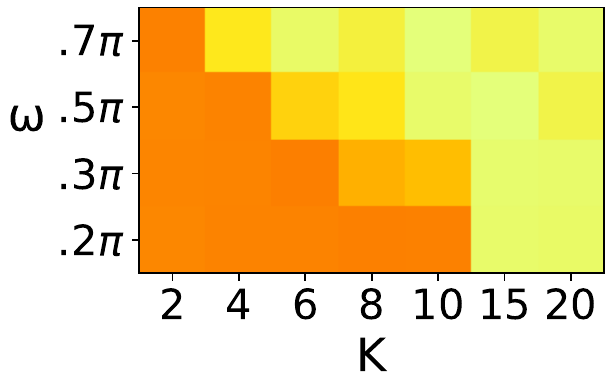}
    }
    \subfloat[Citeseer.]{
    \label{fig:citeseer}
    \includegraphics[width=0.45\linewidth]{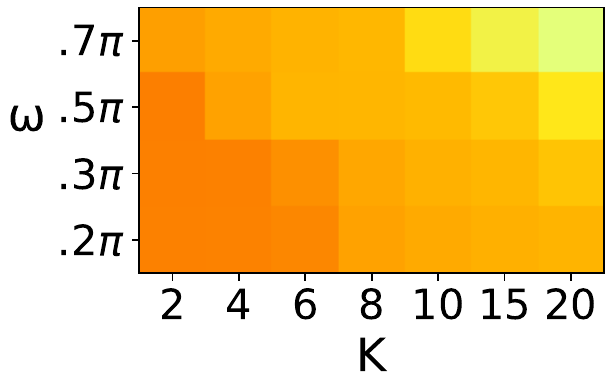}
    }\\ \vskip -0.05in
    \subfloat[Roman.]{
    \label{fig:roman}
    \includegraphics[width=0.45\linewidth]{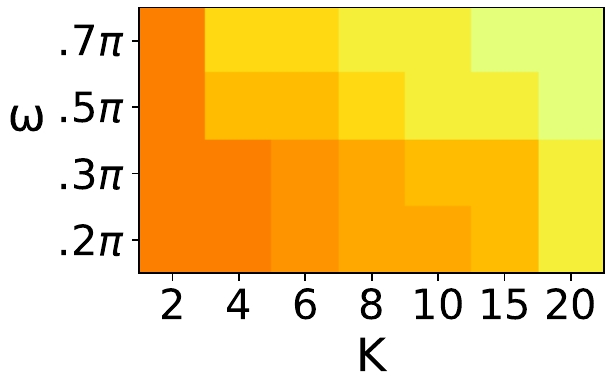}
    }
    \subfloat[Amazon.]{
    \label{fig:amazon}
    \includegraphics[width=0.45\linewidth]{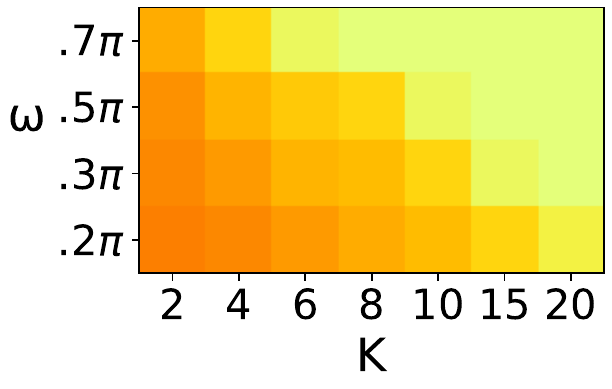}
    }
    \vskip -0.05in
    \caption{Ablation studies on $K$ and $\omega$. 
    Darker shades indicate higher results. 
    Additional results are in Appendix.}
    \label{fig:ablation-K-omega}
\vskip -0.1in
\end{figure}
\begin{table}[!t]
\caption{Ablation studies on Taylor expansion degree.}
\vskip -0.1in
\label{table-ablation-maximum-degree}
\centering
\setlength{\tabcolsep}{4pt}
\resizebox{\linewidth}{!}{
\begin{tabular}{cccccc}
\hline
Degree & $5$                & $10$               & $15$               & $20$               & $25$               \\ \hline
Cora & $88.66_{\pm0.3}$ & $89.21_{\pm0.4}$ & $89.15_{\pm0.2}$ & $89.53_{\pm0.3}$ & $89.28_{\pm0.3}$ \\
Arxiv  & $73.14_{\pm0.2}$ & $75.23_{\pm0.2}$ & $74.74_{\pm0.2}$ & $75.06_{\pm0.2}$ & $74.92_{\pm0.2}$ \\
Roman. & $72.67_{\pm1.0}$ & $74.94_{\pm1.1}$ & $74.83_{\pm1.1}$ & $74.92_{\pm1.2}$ & $75.02_{\pm1.1}$ \\
Genius & $90.02_{\pm0.3}$ & $92.40_{\pm0.2}$ & $91.88_{\pm0.3}$ & $91.83_{\pm0.2}$ & $92.05_{\pm0.3}$ \\ 
\hline
\end{tabular}}
\vskip -0.05in
\end{table}
\begin{table*}[!t]
\caption{Graph anomaly detection results. $^{\ddag}$ Improvements are relative to general-purpose methods rather than GAD baselines.}
\vskip -0.1in
\label{table-GAnoDet}
\centering
\begin{tabular}{cc|cc|cc|cc}
\hline
\multirow{2}{*}{Type}                               & Dateset                                           & \multicolumn{2}{c|}{YelpChi ($1\%$)}              & \multicolumn{2}{c|}{Amazon ($1\%$)}               & \multicolumn{2}{c}{T-Finance ($1\%$)}             \\
                                                      & Metric                                            & F1-macro                & AUROC                   & F1-macro                & AUROC                   & F1-macro                & AUROC                   \\ \hline
\multirow{3}{*}{GAD Models}                           & PC-GNN                                            & $60.55$                 & $75.29$                 & $82.62$                 & $\underline{91.61}$     & $83.40$                 & \textbf{91.85}    \\
                                                      & CARE-GNN                                          & $61.68$                 & $73.95$                 & $75.78$                 & $88.79$                 & $\underline{86.03}$     & $91.17$                 \\
                                                      & GDN                                               & $\underline{65.72}$                 & $75.33$                 & $\underline{90.49}$     & \textbf{92.07}    & $77.38$                 & $89.42$                 \\ \hline
\multirow{2}{*}{\makecell[c]{GAD-specialized \\ spectral GNNs}} & BWGNN                                             & \textbf{66.52}   & $\underline{77.23}$     & $90.28$                 & $89.19$                 & $85.56$                 & $91.38$                 \\
                                                      & GHRN                                              & $62.77$                 & $74.64$                 & $86.65$                 & $87.09$                 & $80.70$                 & $\underline{91.55}$                 \\ \hline
\multirow{5}{*}{\makecell[c]{General-purpose \\ spectral GNNs}}       & GCN                                               & $50.66$                 & $54.31$                 & $69.79$                 & $85.18$                 & $75.26$                 & $87.05$                 \\
                                                      & GPRGNN                                            & $60.45$                 & $67.44$                 & $83.71$                 & $85.28$                 & $77.53$                 & $85.69$                 \\
                                                      & OptBasis                                          & $62.03$                 & $68.32$                 & $86.12$                 & $85.02$                 & $79.28$                 & $86.22$                 \\
                                                      & AdaptKry                                          & $63.40$                 & $66.18$                 & $83.30$                 & $84.58$                 & $80.67$                 & $85.41$                 \\
                                                      & NFGNN                                             & $60.66$                 & $67.36$                 & $85.61$                 & $86.88$                 & $82.38$                 & $86.59$                 \\ \hline
\multirow{2}{*}{\textbf{Ours}}    & TFGNN         & $65.60$     & \textbf{78.79}    &  \textbf{91.10}
                                                                    & $90.12$     & \textbf{87.02}   & $91.42$     \\
                                                      & \#Improv.$^{\ddag}$ & $2.20\%$ & $10.47\%$ & $4.98\%$ & $3.24\%$ & $4.64\%$ & $4.37\%$ \\ \hline
\end{tabular}
\vskip -0.05in
\end{table*}

\subsubsection{\bf Main results and discussions}
\label{section-exp-node-classify-main-results}\

\subsubsection*{\bf Effectiveness of TFGNN} Our TFGNN achieves remarkable advancements in performance on both heterophilic and homophilic graphs. 
Specifically, across all 13 datasets, TFGNN not only leads in performance but does so with improvements of up to \textbf{6.55} over the closest competitor on the Snap-patents dataset.

Furthermore, the advantages of TFGNN are significantly more pronounced when evaluated on heterophilic datasets. 
This trend is corroborated by the numerical findings in Table~\ref{table-numerical-trigo}, which reveal TFGNN’s enhanced capacity to construct functions that can accommodate complex patterns.
Existing studies have empirically shown that heterophilic graphs generally require significantly more complex target filters than the low-pass filters used for homophilic graphs~\cite{BernNet-GNN-narrowbandresults-1,decoupled-NewtonNet,decoupled-AdaptKry}. 
While these complex functions can complicate performance for other methods, TFGNN utilizes its advanced trigonometric filters to navigate these challenges, yielding substantial improvements on heterophilic scenarios.

\subsubsection*{\bf Scalability and Efficiency} Table~\ref{table-node-classify-large} presents a comparative analysis of our TFGNN method alongside leading counterparts, with each baseline recognized for its exceptional scalability and efficiency on large graphs. 
Notably, TFGNN demonstrates superior performance, significantly exceeding all baselines across every dataset while maintaining efficiency comparable to the top-performing methods. 
These findings align with our expectations, as the model complexity—both in terms of computation and parameters—of TFGNN is on par with that of other approaches, as detailed in Section~\ref{section-TFGNN-complexity}.

\subsubsection{\bf Ablation studies}
\label{section-exp-node-classify-ablation}\ 

\subsubsection*{\bf Ablation studies on $K$ and $\omega$} We conduct ablation studies on the two pivotal hyperparameters, $K$ and $\omega$, associated with our trigonometric filters. 
Partial results are illustrated in Figure~\ref{fig:ablation-K-omega}, while a more comprehensive analysis can be found in Appendix.

The results reveal a notable trend: for all datasets, the optimal values for $K$ and $\omega$ tend to fall within low ranges, specifically $K\in\{2,4,6\}$ and $\omega\in\{0.2\pi,0.3\pi,0.5\pi\}$. 
Furthermore, their product $K\cdot\omega$ consistently converges to a similar range across all datasets, approximately $K\cdot\omega\in(0.6\pi,1.2\pi)$. 
This finding aligns with Theorem~\ref{theorem:finiteterms}, which indicates that high-degree terms contribute unnecessary complexity. 
We thus recommend initializing $K$, $\omega$, and $K\cdot\omega$ within these ranges for efficient use of our models, with further fine-tuning as needed for performance optimization.

\subsubsection*{\bf Ablation studies on $D$} We perform ablation studies on the degree of Taylor expansion, $D$. 
Table~\ref{table-ablation-maximum-degree} shows that while increasing the degree improves performance to a certain extent, accuracy eventually stabilizes. 
This is consistent with prior studies and can be understood in terms of polynomial approximation. 
Higher-degree orthogonal bases tend to minimize approximation loss; however, beyond an optimal degree, further increases become negligible~\cite{polyapprox_1,polyapprox_2}.

\subsection{Application on graph anomaly detection}
\label{section-exp-graph-anomaly-detection}

We investigate an application example of TFGNN for the graph anomaly detection (GAD) task, which is typically recognized as binary node classification task (normal vs. abnormal)~\cite{GAnoDet-survey-1,GAnoDet-survey-2}.

\subsubsection{\bf Datasets and baselines}\
\subsubsection*{\bf Datasets.} We adopt three datasets (YelpChi, Amazon, and T-Finance) with a low label-rate of $1\%$ set across all datasets following~\cite{GAnoDet-spectral-1-BWGNN}.

\subsubsection*{\bf Baselines and model implementations.} We include $10$ baseline methods, organized into three types below:
\begin{itemize}[leftmargin=*,parsep=0pt,itemsep=2pt,topsep=2pt]
\item {\bf GAD models:} PC-GNN~\cite{GAnoDet-2-PC-GNN}, CARE-GNN~\cite{GAnoDet-1-CARE-GNN}, GDN~\cite{GAnoDet-3-GDN}.
\item {\bf GAD-specialized spectral GNNs:} BWGNN~\cite{GAnoDet-spectral-1-BWGNN}, GHRN~\cite{GAnoDet-spectral-2-GHRN}.
\item {\bf General-purpose spectral GNNs:} GCN~\cite{GCN}, GPRGNN~\cite{GPRGNN}, OptBasis~\cite{OptBasisGNN}, AdaptKry~\cite{decoupled-AdaptKry}, NFGNN~\cite{decoupled-NFGNN}.
\end{itemize}
The specifications for implementing common baselines (PC-GNN, CARE-GNN, BWGNN, GCN) are derived from~\cite{GAnoDet-spectral-1-BWGNN}. 
In our TFGNN and other general-purpose methods, we utilize a two-layer MLP with $64$ hidden units for the feature transformation module, maintaining alignment with the GAD-specialized models. 
The hyperparameters for TFGNN are optimized as detailed in Section~\ref{section-exp-node-classify}, while the other baselines follow the configurations outlined in their original papers. 
More experimental details are in Appendix.

\subsubsection{\bf Main results and discussions}
\label{section-exp-graph-anomaly-detection-main-results}\ 


\subsubsection*{\bf Improvements on specific task.} 
Table~\ref{table-GAnoDet} highlights the \#Improv. metric, showing that TFGNN outperforms general-purpose models significantly, achieving increases of up to $11.34\%$ on the YelpChi dataset. 
This suggests that while general-purpose spectral GNNs can perform well in benchmark node classification tasks, they often underperform in specialized applications. 
In contrast, TFGNN, with its advanced graph filters, consistently provides notable improvements across both standard and specialized tasks, demonstrating the effectiveness and versatility of our approach.

\subsubsection*{\bf Comparable to GAD-specialized spectral GNNs.} Table~\ref{table-GAnoDet} indicates that TFGNN's performance rivals that of specialized spectral GNNs for GAD. 
Models like BWGNN and GHRN, which are built on the same graph spectrum principles, incorporate specific features aimed at enhancing performance. 
For example, BWGNN~\cite{GAnoDet-spectral-1-BWGNN} effectively addresses the “right-shift” phenomenon with its customized beta wavelets, while GHRN~\cite{GAnoDet-spectral-2-GHRN} focuses on filtering out high-frequency components to prune inter-class edges in heterophilic graphs. 
In contrast, TFGNN offers a unique and effective filtering strategy for GAD tasks, showcasing impressive outcomes. 
This reflects a promising direction for improving spectral GNNs through the introduction of more advanced polynomial graph filters.


\section{Conclusions}
\label{section-conclusion}

In this paper, we address the polynomial selection problem in spectral GNNs, linking polynomial capabilities to their effectiveness. 
We present the first proof that the construction error of graph convolution layers is bounded by the sum of polynomial approximation errors on function slices, supported by intuitive numerical validations. 
This insight motivates the use of ``narrow function-preferred'' polynomials, leading to the introduction of our advanced trigonometric graph filters. 
The proposed filters not only demonstrate provable parameter-efficiency but also employ Taylor-based parameter decomposition for streamlined implementation. 
Building upon this, we introduce TFGNN, a scalable spectral GNN featuring a decoupled architecture. 
The efficacy of TFGNN is confirmed through benchmark node classification tasks and a practical example in graph anomaly detection, highlighting the adaptability and real-world relevance of our theoretical contributions.

\subsubsection*{\bf Limitations and future works.} Our theoretical framework is grounded in the concept of function slices, which are inherently linked to target filters. 
However, in practical scenarios, the diversity and variability of target filters can hinder the specificity of our theoretical results, potentially leading to suboptimal solutions. 
Therefore, a promising future research is to categorize these filters and analyze their numerical properties, thereby enabling more consistent enhancements in the design of spectral GNNs.


\bibliographystyle{ACM-Reference-Format}
\bibliography{references}


\appendix


\section{Proof of Lemma~\ref{lemma:inequality-polynomial-error}}
\label{appendix:proof-to-lemma:inequality-polynomial-error}

\begin{proof}
\label{proof:lemma:inequality-polynomial-error}
We establish this inequality by proving its right-hand and left-hand sides independently.

\subsubsection*{Proof of right hand side} For convenience, we define the L$_2$ norm of a function $g$ over the interval $\left[0,2\right]$, denoted by $\lVert g \rVert_{2}$
 , as follows:
\begin{equation}
\label{eq:proof-lemma-1}
\lVert g \rVert_{2} \triangleq (\int_{0}^{2}\lvert g(x) \rvert^{2}\ dx)^{\frac{1}{2}}
\end{equation}
Using the norm expression defined earlier, and recalling the expression for $\epsilon$ from Eq.~\ref{eq:polynomial-approximation-error-continuous}, we can derive the following inequalities by applying the Cauchy-Schwarz inequality:
\begin{align}
\label{eq:proof-lemma-2}
\epsilon 
&= \lVert f(x) - \mathbf{T}_{0:D}(x;f) \rVert_{2}^{2}\ , \notag \\
&= \lVert \sum_{s=1}^{n}f_{s}(x) - \sum_{s=1}^{n}\mathbf{T}_{0:D}(x;f_{s}) \rVert_{2}^{2}\ , \\
&\leq (\sum_{s=1}^{n}\lVert f_{s}(x) - \mathbf{T}_{0:D}(x;f_{s}) \rVert_{2})^{2} = (\sum_{s=1}^{n}\sqrt{\epsilon_{s}})^{2}\ ,
\end{align}
which is our right-hand side.

\subsubsection*{Proof of Left-hand side} To proceed without loss of generality, we consider $f$ to be nonnegative over the entire interval. 
Recalling the definition of $\epsilon$ from Eq.~\ref{eq:polynomial-approximation-error-continuous}, it follows that
\begin{align}
\label{eq:proof-lemma-3}
\epsilon
&= \lVert\mathbf{T}_{0:D}(x;f)-f(x)\rVert^{2}_{2}\ ,\notag\\
&= \lVert\sum_{s=1}^{n}f_{s}(x) - \sum_{s=1}^{n}\mathbf{T}_{0:D}(x;f_{s})\rVert^{2}_{2}\ ,\\
&=2\sum_{1\leq p \leq q \leq n} \lVert\sqrt{
det\left|
\begin{array}{cc}
   f_{p}(x)  &  f_{q}(x)\\
    \mathbf{T}_{0:D}(x;f_{q}) & \mathbf{T}_{0:D}(x;f_{p})
\end{array}
\right|
}\rVert^{2}_{2} \notag\\
&\quad\quad\quad +\sum_{s=1}^{n}\sqrt{\epsilon_{s}}^{2}\ ,\\
&\geq 0 + \sum_{s=1}^{n}\sqrt{\epsilon_{s}}^{2} = \sum_{s=1}^{n}\epsilon_{s}\ ,
\end{align}
which is our left-hand side.

Thus, combining the two parts of the proof above, we confirm that Lemma~\ref{lemma:inequality-polynomial-error} holds for the continuous form of error.
\end{proof}

\subsubsection*{\bf Adaptation to the disctrete error form} This is due to the applicability of the Cauchy-Schwarz inequality to the discrete version of norm inequalities, ensuring that the right-hand side holds for the discrete form of the error. 
The left-hand side, which is based solely on fundamental non-negative relations, also maintains the inequality in the discrete setting. 
Consequently, Lemma~\ref{lemma:inequality-polynomial-error} can be directly adapted to the discrete scenario.


\section{Proof of Theorem~\ref{theorem:inequality-gnn-layer-error}}
\label{appendix:proof-to-theorem:inequality-gnn-layer-error}

\begin{proof}
\label{proof:theorem:inequality-gnn-layer-error}
We establish this inequality by proving its right-hand and left-hand sides independently.

\subsubsection*{Proof of right hand side} Recalling the expression of $\xi$ from Eq.~\ref{eq:graph-convolution-layer-construction-error}, we can derive the following inequality using the submultiplicative property of Frobenius norm:
\begin{align}
\label{eq:proof-theorem-gnn-error-1}
\xi=&\lVert\boldsymbol{U}diag(\mathbf{T}_{0:D}(\boldsymbol{\lambda};f)-f(\boldsymbol{\lambda}))\boldsymbol{U}^{T}\boldsymbol{X}\boldsymbol{W}\rVert_{F} \notag\\
\leq&\lVert\boldsymbol{U}diag(\mathbf{T}_{0:D}(\boldsymbol{\lambda};f)-f(\boldsymbol{\lambda}))\boldsymbol{U}^{T}\rVert_{F}\cdot\lVert\boldsymbol{X}\rVert_{F}\cdot\lVert\boldsymbol{W}\rVert_{F}\ ,\\
\leq&r\cdot\lVert\boldsymbol{U}diag(\mathbf{T}_{0:D}(\boldsymbol{\lambda};f)-f(\boldsymbol{\lambda}))\boldsymbol{U}^{T}\rVert_{F}\cdot\lVert\boldsymbol{X}\rVert_{F}\ .
\end{align}
Note that $\boldsymbol{U}$ is orthogonal matrix, which will not influence the Frobenius norm of any matrices in product operation. 
Thus, the inequality above can further be derived as:
\begin{align}
\label{eq:proof-theorem-gnn-error-2}
\xi\leq&r\cdot\lVert\boldsymbol{U}diag(\mathbf{T}_{0:D}(\boldsymbol{\lambda};f)-f(\boldsymbol{\lambda}))\boldsymbol{U}^{T}\rVert_{F}\cdot\lVert\boldsymbol{X}\rVert_{F}\ \notag\\
=&r\cdot\lVert diag(\mathbf{T}_{0:D}(\boldsymbol{\lambda};f)-f(\boldsymbol{\lambda}))\rVert_{F}\cdot\lVert\boldsymbol{X}\rVert_{F}\ ,\\
=&r\cdot \epsilon \cdot \lVert\boldsymbol{X}\rVert_{F}\ ,\\
\leq&r\lVert\boldsymbol{X}\rVert_{F}(\sum_{s=1}^{n}\sqrt{\epsilon_{s}})^{2}\ ,
\end{align}
which is the right-hand side.

\subsubsection*{Proof of left hand side} Using the basic of the matrix perturbation theory, we can derive the following inequality:
\begin{align}
\label{eq:proof-theorem-gnn-error-3}
\xi=&\lVert\boldsymbol{U}diag(\mathbf{T}_{0:D}(\boldsymbol{\lambda};f)-f(\boldsymbol{\lambda}))\boldsymbol{U}^{T}\boldsymbol{X}\boldsymbol{W}\rVert_{F} \notag\\
\geq&\delta_{\boldsymbol{X}}\delta_{\boldsymbol{W}}\lVert\boldsymbol{U}diag(\mathbf{T}_{0:D}(\boldsymbol{\lambda};f)-f(\boldsymbol{\lambda}))\boldsymbol{U}^{T}\rVert_{F}\ ,\\
\geq&\delta_{\boldsymbol{X}}\delta_{\boldsymbol{W}}\sum_{s=1}^{n}\epsilon_{s}\ ,
\end{align}
which is the left-hand side.

Thus, combining the two parts of the proof above, we confirm that Theorem~\ref{theorem:inequality-gnn-layer-error} holds.
\end{proof}


\section{Proof of Theorem~\ref{theorem:finiteterms}}
\label{appendix:proof-to-theorem:finiteterms}

\begin{proof}
\label{proof:finiteterms}
To begin with, note that $\alpha_{k}$ and $\beta_{k}$ can be computed as follows:
\begin{align}
\label{comp-walpha-wbeta-1}
\alpha_{k} &= \frac{\omega}{\pi}\int_{0}^{\frac{2\pi}{\omega}}f(x)\sin(k\omega x)\,dx \ , \notag \\  
\beta_{k} &= \frac{\omega}{\pi}\int_{0}^{\frac{2\pi}{\omega}}f(x)\cos(k\omega x)\,dx \ .
\end{align}
We alternatively consider such a \textit{real function} $f'(x)$ which is defined as follows:
\begin{equation}
\label{tilde-f}
f'(x)=\begin{cases}
f(x) \ , \quad &x\in\left[ 0,\frac{2\pi}{\omega}\right] \ ; \\
0 \ , \quad &others \ .
\end{cases}
\end{equation}
Notice that such $f'(x)$ satisfies the \textit{Dirichlet conditions}~\cite{deltafunc}, the \textit{Fourier transform} of $f'(x)$ exists according to~\cite{Digitalsignalprocessing} and is defined as follows:
\begin{align}
\label{FT-tilde-f-1}
F'(\Omega)&=\int_{-\infty}^{+\infty}f'(x)e^{-i\Omega x}\, dx \ , \notag \\
          &=\int_{-\infty}^{+\infty}f'(x)\lbrack\cos(\Omega x)-i\cdot\sin(\Omega x)\rbrack\, dx \ ,
\end{align}
where $\Omega$ denotes the variable in frequency domain, and $i$ is the imaginary unit. 
\par Furthermore, the $f'\in L^{1}(\mathbb{R}^{n})$ is an integrable function, making $f'(x)$ satisfy the \textit{Riemann–Lebesgue lemma} which is defined as follows:
\begin{theorem}
\label{riem-lebes}\textit{(\textbf{Riemann–Lebesgue lemma~\cite{RiemLebg}}). 
Let $g\in L^{1}(\mathbb{R}^{n})$ be an integrable function, and let $G$ be the Fourier transform of $g$. 
Then the $G$ vanishes at infinity, which is defined as follows:}
\begin{equation}
\label{eq-riem-lebes}
\lim_{\left|\Omega\right|\to \infty}\left|G(\Omega)\right|=0 \ .
\end{equation}
\end{theorem}
Thus, the limit of $F'(\Omega)$ as $\Omega$ approaches $+\infty$ equals $0$, which is defined as
\begin{align}
\label{FT-tilde-f-2}
\lim_{\Omega\to +\infty}F'(\Omega)&=\underbrace{\lim_{\Omega\to +\infty}\left[\int_{-\infty}^{+\infty}f'(x)\lbrack\cos(\Omega x)-i\cdot\sin(\Omega x)\rbrack\, dx\right]}_{\text{Formula 1}} \ , \notag \\
                                  &=0 \ .
\end{align}
Since the $f'(x)$ is a real function, the Formula 1 equals $0$ if and only if the following equations hold:
\begin{align}
\label{FT-tilde-f-3}
&\lim_{\Omega\to +\infty}\int_{-\infty}^{+\infty}f'(x)\cos(\Omega x)\, dx=0 \ , \notag \\
&\lim_{\Omega\to +\infty}\int_{-\infty}^{+\infty}f'(x)\sin(\Omega x)\, dx=0 \ .
\end{align}
With combing the definition of $f'(x)$ in Eq. \ref{tilde-f}, the Eq. \ref{FT-tilde-f-3} above can be further derived as follows:
\begin{align}
\label{FT-tilde-f-4}
&\lim_{\Omega\to +\infty}\int_{0}^{\frac{2\pi}{\omega}}f(x)\cos(\Omega x)\, dx=0 \ , \notag \\
&\lim_{\Omega\to +\infty}\int_{0}^{\frac{2\pi}{\omega}}f(x)\sin(\Omega x)\, dx=0 \ .
\end{align}
Finally, with replacing the $\Omega$, $\Omega\to +\infty$ with $k\omega$, $k\to +\infty$ in Eq. \ref{FT-tilde-f-4}, and further considering the Eq. \ref{comp-walpha-wbeta-1}, we obtain the following equations:
\begin{align}
\label{comp-walpha-wbeta-2}
\lim_{k\to +\infty}\alpha_{k}=\frac{\omega}{\pi}\cdot\lim_{k\to +\infty}\int_{0}^{\frac{2\pi}{\omega}}f(x)\sin(k\omega x)\,dx=0 \ , \notag \\
\lim_{k\to +\infty}\beta_{k}=\frac{\omega}{\pi}\cdot\lim_{k\to +\infty}\int_{0}^{\frac{2\pi}{\omega}}f(x)\cos(k\omega x)\,dx=0 \ ,
\end{align}
and the proof is completed. 
\end{proof}





\section{Related Works}
\label{appendx:related-works}

\subsection{Spectral-based graph neural networks}
\label{appendx:related-works-spectral-GNNs}
Spectral-based graph neural networks form a unique branch of GNNs designed to process graph-structured data by applying graph filters to execute graph convolution (filtering) operations~\cite{surveyspectralgnn}. The pioneering spectral GNN, SpectralCNN\cite{spectralCNN}, was developed as a generalization of convolutional neural networks for graph data, using principles from spectral graph theory. 
Subsequent refinements, such as ChebNet\cite{ChebNet} and GCN~\cite{GCN}, have built upon this foundation.

In recent advancements, the design of spectral GNNs has increasingly focused on incorporating various graph filters, which are central to their functionality. 
Polynomial approximation has become the prevailing approach for constructing these filters, providing both enhanced performance and operational efficiency. 
As a result, many contemporary spectral GNNs are predominantly defined by polynomial frameworks. 
For instance, GPRGNN~\cite{GPRGNN} introduces a monomial-based graph filter, interpreted as a generalized PageRank algorithm. 
BernNet~\cite{ChebNet} leverages Bernstein polynomials to create nonnegative graph filters, demonstrating significant effectiveness in real-world applications. 
JacobiConv~\cite{JacobiConv} unifies different methods by employing Jacobian polynomials.
OptBasis~\cite{OptBasisGNN} improves the design of spectral GNNs by introducing filters with optimal polynomial bases. 
UniFilter~\cite{decoupled-UniFilter} introduces the notion of universal bases, bridging polynomial filters with graph heterophily.

\subsection{Node classification with heterophily}
\label{appendx:related-works-node-classify-heterophily}
In recent years, heterophilic graphs have drawn considerable interest in the field of graph learning. 
Unlike traditional homophilic graphs, where linked nodes usually share the same label, heterophilic graphs connect nodes with contrasting labels. 
This unique structure presents significant challenges for graph neural networks (GNNs)\cite{heterophily-gnn-survey,heterophily-gnn-survey-2,heterophily-gnn-survey-3}, which are typically designed for homophilic settings. 
To address these challenges, a range of GNNs tailored to heterophily have emerged. 
For example, H2GCN\cite{H2GCN} introduces specialized mechanisms for embedding nodes in heterophilic environments, OrderGNN~\cite{OrderedGNN} restructures message-passing to account for heterophily, and LRGNN~\cite{low-rank-gnn-2} leverages a global label relationship matrix to improve performance under heterophily.

\subsubsection*{\bf Addressing heterophily with spectral GNNs} In most recent, spectral-based GNNs have shown promise in addressing these challenges by learning dataset-specific filters that extend beyond the standard low-pass filters used in conventional GNNs. 
By doing so, spectral GNNs demonstrate improved performance in tackling heterophilic graphs, achieving superior results in node classification under heterophily~\cite{GPRGNN,BernNet-GNN-narrowbandresults-1,JacobiConv,chebnet2d,ChebNetII,OptBasisGNN,decoupled-UniFilter}.

\subsection{Graph anomaly detection}
\label{appendx:related-works-graph-anomaly}

Graph-based anomaly detection (GAD) is a specialized task within anomaly detection, aimed at identifying anomalies within graph-structured data~\cite{GAnoDet-survey-1, GAnoDet-survey-2}. 
The primary goal in GAD is to detect anomalous nodes (outliers) in the graph by leveraging a limited set of labeled samples, including both anomalous and normal nodes. 
Effectively, GAD can be viewed as a binary node classification task, where the classes represent anomaly and normalcy. 
The recent success of Graph Neural Networks (GNNs) in node classification has spurred the development of GAD-specialized GNN methods, such as CARE-GNN~\cite{GAnoDet-1-CARE-GNN}, PC-GNN~\cite{GAnoDet-2-PC-GNN}, and GDN~\cite{GAnoDet-3-GDN}, with each significantly enhancing detection performance.

\subsubsection*{\bf Spectral GNNs in GAD} Building on the success of GNN-based approaches for graph anomaly detection (GAD), recent studies leveraging spectral GNNs have yielded promising results. 
By framing GAD through graph spectrum analysis, these methods introduce novel perspectives on the problem. 
For example, BWGNN~\cite{GAnoDet-spectral-1-BWGNN} utilizes beta graph wavelets for signal filtering, effectively addressing the ``right-shift'' phenomenon in GAD. 
Similarly, GHRN~\cite{GAnoDet-spectral-2-GHRN} enhances GAD by pruning inter-class edges, focusing on high-frequency graph components to improve detection performance.


\section{Experimental Details}
\label{appendix:exp-details}

This section outlines the extensive experimental settings relevant to the studies conducted in Section~\ref{section-connect-polynomial-ability-spectral-GNN-ability-exp} and Section~\ref{section-exp}. 
Experiments are conducted using an NVIDIA Tesla V100 GPU with 32GB of memory, running on Ubuntu 20.04 OS and CUDA version 11.8.

\subsection{Experimental details of numerical validation}
\label{appendix:exp-details-numerical}

\subsubsection*{\bf Descriptions to target functions} The six target functions employed in the numerical experiments are defined by the expressions presented in Table~\ref{table-datasets-expressions}.
\begin{table*}[!ht]
  \caption{Mathematical expressions of six target functions.} 
  \label{table-datasets-expressions}
  \centering
\begin{tabular}{cc}
\hline
Functions  & Expressions                                                 \\ \hline
$f_{1}(x)$ & $e^{-20(x-0.5)^2} + e^{-20(x-1.5)^2}$                       \\
$f_{2}(x)$ &  $\left\{
\begin{array}{cc}
  e^{-100(x-0.8)^2} + e^{-100(x-1.2)^2} + 0.5 \cdot (1 + \cos(2\pi x))\ ,  & x\in\left[0,0.5\right] \\
  e^{-100(x-0.8)^2} + e^{-100(x-1.2)^2}\ , & x\in\left(0.5,1.5\right) \\
  e^{-100(x-0.8)^2} + e^{-100(x-1.2)^2} + 0.5 \cdot (1 + \cos(2\pi x))\ , & x\in\left[1.5,2\right]
\end{array}
\right.$                                                           \\
$f_{3}(x)$ & $e^{-100(x-0.5)^2} + e^{-100(x-1.5)^2} + 1.5e^{-50(x-1)^2}$ \\
$f_{4}(x)$ & $e^{-100x^2}+e^{-100(x-2)^2}$                               \\
$f_{5}(x)$ & $1-e^{-10x^2}$                                              \\
$f_{6}(x)$ & $e^{-10(x-0.4)^2}+2e^{-10(x-1.5)^2}$                        \\ \hline
\end{tabular}
\end{table*}

\subsubsection*{\bf Random graph construction} We construct random graphs using the \textit{Erdős-Rényi} model, specifically denoted as $G(n, p)$~\cite{NetworksAnIntroduction}. In our experiments, we set the number of nodes $n$ to 50,000 and the edge creation probability $p$ to 0.5.

\subsubsection*{\bf Random node feature construction} We construct random node features $\boldsymbol{X}$ drawn from a Gaussian distribution. Each entry in the feature matrix $\boldsymbol{X}$ is independently sampled and follows a standard normal distribution, $N(0,1)$.

\subsubsection*{\bf Experimental settings} To start, we randomly generate ten pairs of graphs and features, denoted as $(\mathcal{G}_{1},\boldsymbol{X}_{1})$, $(\mathcal{G}_{2},\boldsymbol{X}_{2})$, 
..., $(\mathcal{G}_{10},\boldsymbol{X}_{10})$. 
For each pair $(\mathcal{G}_{j},\boldsymbol{X}_{j})$, we apply six different graph filters, resulting in six filtered outputs: $\boldsymbol{Y}_{1j}$, $\boldsymbol{Y}_{2j}$, ..., $\boldsymbol{Y}_{6j}$. This process involves performing graph convolution on $\boldsymbol{X}$ using these target functions.

As discussed in Section~\ref{section-connect-polynomial-ability-spectral-GNN-ability-exp}, we pursue two tasks: the first task involves approximating function slices, while the second focuses on filter learning.
\begin{itemize}[leftmargin=*,parsep=2pt,itemsep=2pt,topsep=2pt]
\item {\bf Approximation of function slices.} We generate $50000$ function slices for each target function based on the eigenvalues of the graph $\mathcal{G}_{j}$
 . Various polynomial bases are employed for the approximation, with the minimum sum of squared errors (SSE) serving as the evaluation metric. The final results are averaged over ten randomly generated graphs.
\item {\bf Graph filter learning.} We implement a one-layer linear spectral Graph Neural Network (GNN) that operates without a weight matrix $\boldsymbol{W}$, utilizing various polynomial bases. The input consists of pairs $(\mathcal{G}_{j},\boldsymbol{X}_{j})$to approximate the target output $\boldsymbol{Y}$. The learned graph filters are then employed to compute the discrepancies with the target filters, using the Frobenius norm as the evaluation metric. Finally, the results are averaged across ten randomly generated graphs to ensure robustness.
\end{itemize}

\subsection{Experimental details for node classification}
\label{appendix:exp-details-node-classify}

\subsubsection*{\bf Dataset statistics} The statistics of the $13$ datasets used in Section~\ref{section-exp-node-classify} are provided in Tables~\ref{table-datasets-statistics-medium-to-large} and~\ref{table-datasets-statistics-exlarge}. 

\begin{table*}[!ht]
  \caption{Statistics for medium-to-large datasets, with \# Edge homo indicating the edge homophily measure from~\cite{H2GCN}.} 
  \vskip -0.05in
  \label{table-datasets-statistics-medium-to-large}
  \centering
  \begin{tabular}{lccccccccc}
  \hline
    & Cora & CiteSeer & PubMed & Ogbn-arxiv & Roman-empire & Amazon-ratings & Questions & Gamers & Genius \\ \hline
    \# Nodes & 2708 & 3327 & 19,717 &  169,343  &  22,662  &  24,492  &  48,921 & 168,114  &  \textbf{421,961}  \\
    \# Edges &  5278 & 4552 & 44,324 & 1,157,799 & 32,927 & 93,050 & 153,540 & \textbf{6,797,557} & 922,868 \\
    \# Features & 1433 & 3703 & 500 & 128 & 300 & 300 & 301 & 7 & 12 \\
    \# Classes & 7 & 6 & 5 & 40 & 18 & 5 & 2 & 2 & 2 \\
    \# Edge homo~\cite{H2GCN} &  0.81 &  0.74 &  0.80 &  0.65 &  0.05 &  0.38 & 0.84 &  0.55 &  0.62 \\
    \hline
  \end{tabular}
  \vskip -0.05in
\end{table*}
\begin{table*}[!ht]
  \caption{Statistics for exceptionally large datasets. 
  \# Edge homo for Ogbn-papers100M is unavailable due to runtime exceedance.} 
  \vskip -0.05in
  \label{table-datasets-statistics-exlarge}
  \centering
  \begin{tabular}{lcccc}
  \hline
    & Ogbn-products & Ogbn-papers100M & Snap-patents & Pokec \\ \hline
    \# Nodes &  2,449,029 & \textbf{111,059,956} & 2,923,922 &  1,632,803 \\
    \# Edges &   61,859,140 & \textbf{1,615,685,872} & 13,975,788 & 30,622,564 \\
    \# Features & 100 & 128 & 269 & 65 \\
    \# Classes & 47 & 172 & 5 & 2 \\
    \# Edge homo~\cite{H2GCN} &  0.81 &  - & 0.07 &  0.45  \\
    \hline
  \end{tabular}
  \vskip -0.05in
\end{table*}

\subsubsection*{\bf Baseline implementations} We provide code URLs to the public implementations for all baselines referenced in this paper. 
In particular, for the well-established baselines GCN and ChebNet, we employ standardized implementations based on previous research~\cite{BernNet-GNN-narrowbandresults-1,ChebNetII,chebnet2d,OptBasisGNN,JacobiConv,decoupled-FEGNN,decoupled-AdaptKry,decoupled-NFGNN}; 
for the remaining baselines, we resort to the publicly released code, accessible via the provided URLs as below.

\begin{itemize}[leftmargin=*,parsep=2pt,itemsep=2pt,topsep=2pt]
\item H2GCN: \url{https://github.com/GemsLab/H2GCN}
\item GloGNN: \url{https://github.com/RecklessRonan/GloGNN}
\item LINKX: \url{https://github.com/CUAI/Non-Homophily-Large-Scale}
\item OrderGNN: \url{https://github.com/lumia-group/orderedgnn}
\item LRGNN: \url{https://github.com/Jinx-byebye/LRGNN}
\item GCN: \url{https://github.com/ivam-he/ChebNetII}
\item SGC: \url{https://github.com/ivam-he/ChebNetII}
\item GCNII: \url{https://github.com/chennnM/GCNII}
\item ChebNet: \url{https://github.com/ivam-he/ChebNetII}
\item ACMGCN: \url{https://github.com/SitaoLuan/ACM-GNN}
\item Specformer: \url{https://github.com/DSL-Lab/Specformer}
\item GPRGNN: \url{https://github.com/jianhao2016/GPRGNN}
\item BernNet: \url{https://github.com/ivam-he/BernNet}
\item ChebNetII: \url{https://github.com/ivam-he/ChebNetII}
\item OptBasis: \url{https://github.com/yuziGuo/FarOptBasis}
\item NFGNN: \url{https://github.com/SsGood/NFGNN}
\item JacobiConv: \url{https://github.com/GraphPKU/JacobiConv}
\item AdaptKry: \url{https://github.com/kkhuang81/AdaptKry}
\item UniFilter: \url{https://github.com/kkhuang81/UniFilter}
\end{itemize}

\subsubsection*{\bf Implementation of TFGNN} 

As introduced in Section~\ref{section-method-decoupled}, TFGNN is implemented in two distinct configurations to accommodate graphs of varying sizes. 
For graphs detailed in Table~\ref{table-node-classify-medium}, we utilize the architecture represented by Eq.~\eqref{eq:decoupled-TFGNN-medium}. 
For larger graphs listed in Table~\ref{table-node-classify-large}, we employ the architecture shown in Eq.~\eqref{eq:decoupled-TFGNN-large}.

The MLP architecture within TFGNN is dataset-specific. 
For medium-sized graphs (Cora, Citeseer, Pubmed, Roman-empire, Amazon-ratings, and Questions), we use a two-layer MLP with $64$ hidden units. 
In contrast, larger datasets are assigned three-layer MLPs with varying hidden units: $128$ for Gamers and Genius, $256$ for Snap-patents and Pokec, $512$ for Ogbn-arxiv, and $1024$ for Ogbn-papers100M.

To ensure experimental fairness, we fix the order of the Taylor-based parameter decomposition, denoted as $D$, to $10$, aligning with other baselines such as GPRGNN and ChebNetII. 
We employ a grid search to optimize the weight decay over $\{5e-1,5e-2,5e-3,5e-4,0\}$, learning rate over $\{0.5,0.1,0.05,0.01,0.005,0.001\}$, dropout over $\{0,0.2,0.5,0.7,0.9\}$, $\omega$ within $\{0.2\pi,0.3\pi,0.5\pi,0.7\pi\}$, and $K$ from $\{2,4,6,8,10,15,20\}$.

\subsubsection*{\bf Model training and testing} We follow the dataset splitting protocols established in the literature. 
For the Cora, Citeseer, and Pubmed datasets, we utilize the established $60\%/20\%/20\%$ train/val/test split, which has been widely adopted across numerous studies~\cite{BernNet-GNN-narrowbandresults-1,ChebNetII,chebnet2d,JacobiConv,OptBasisGNN,decoupled-FEGNN,decoupled-PCConv,decoupled-UniFilter,decoupled-AdaptKry}. 
For the Roman-empire, Amazon-ratings, and Questions datasets, we implement a $50\%/25\%/25\%$ train/val/test split, aligning with the protocols outlined in their original publications~\cite{dataset8-small-hetero}. 
This $50\%/25\%/25\%$ train/val/test split strategy is also applied to the Gamers, Genius, Snap-patents, and Pokec datasets, as recommended in~\cite{dataset6-large-hetero}. 
Finally, for Ogbn-arxiv, Ogbn-products, and Ogbn-papers100M, we adopt the fixed splits defined in the original OGB dataset paper~\cite{dataset5-ogb}.

Models are trained for a maximum of $1,000$ epochs, with early stopping implemented after $200$ epochs if there is no improvement in validation accuracy. 
To handle exceptionally large graphs, we employ a mini-batch training strategy using batches of $20,000$ nodes. 
The optimization process employs the Adam optimizer~\cite{Adamoptimizer}. 
For each dataset, we generate $10$ random node splits and perform $10$ random initializations for each baseline on these splits. 
This process yields a total of $100$ evaluations for each dataset. 
The reported results for each baseline represent the average of these $100$ evaluations.

\subsection{Experimental details for graph anomaly detection}
\label{appendix:exp-details-graph-anomaly-detection}

\subsubsection*{\bf Dataset statistics} Table~\ref{table-datasets-statistics-GAnodet} presents the statistics of datasets used in Section~\ref{section-exp-graph-anomaly-detection}. 

\begin{table}[!ht]
  \caption{Statistics of datasets utilized for graph anomaly detection. 
  \# Anomaly represents the rate of abnormal nodes.}
  \vskip -0.05in
  \label{table-datasets-statistics-GAnodet}
  \centering
  \begin{tabular}{lccc}
  \hline
    & YelpChi & Amazon & T-Finance \\ \hline
    \# Nodes &  45,954  & 11,944 & 39,357  \\
    \# Edges &  3,846,979 & 4,398,392 & 21,222,543  \\
    \# Features & 32 & 25 & 10 \\
    \# Anomaly & 14.53\% & 6.87\% &  4.58\% \\
    \hline
  \end{tabular}
  \vskip -0.05in
\end{table}

\subsubsection*{\bf Baseline implementations} We provide code URLs to the official implementations of all baseline models referenced in this paper. 
Specifically, for general-purpose spectral GNNs like GPRGNN, OptBasis, AdaptKry, and NFGNN, which are initially introduced as uniform, decoupled GNN architectures, we implement them in alignment with the TFGNN variant defined in Eq.~\eqref{eq:decoupled-TFGNN-medium}. 
Each model uses a fixed maximum polynomial degree of $10$ and a two-layer MLP with $64$ hidden units for feature transformation, consistent with BWGNN~\cite{GAnoDet-spectral-1-BWGNN}. 
The GCN baseline is similarly implemented with a two-layer setup featuring $64$ hidden dimensions. 
For other baselines, we rely on their official implementations (links provided below). 
All models are rebuilt and evaluated in PyG~\cite{PyTorchGeometric} framework to maintain experimental fairness.

\begin{itemize}[leftmargin=*,parsep=2pt,itemsep=2pt,topsep=2pt]
\item PC-GNN: \url{https://github.com/PonderLY/PC-GNN}
\item CARE-GNN: \url{https://github.com/YingtongDou/CARE-GNN}
\item GDN: \url{https://github.com/blacksingular/wsdm_GDN}
\item BWGNN: \url{https://github.com/squareRoot3/Rethinking-Anomaly-Detection}
\item GHRN: \url{https://github.com/blacksingular/GHRN}
\item GPRGNN: \url{https://github.com/jianhao2016/GPRGNN}
\item OptBasis: \url{https://github.com/yuziGuo/FarOptBasis}
\item AdaptKry: \url{https://github.com/kkhuang81/AdaptKry}
\item NFGNN: \url{https://github.com/SsGood/NFGNN}
\end{itemize}

\subsubsection*{\bf Implementation of TFGNN} 
In pursuit of fairness, TFGNN incorporates a decoupled architecture consistent with general-purpose spectral GNNs, featuring a maximum polynomial degree of 10 and a two-layer MLP comprising 64 hidden units for feature transformation. 
This approach also ensures that parameter fairness is in alignment with BWGNN. 
For hyperparameter tuning, we adhere to the previous setups detailed in Appendix~\ref{appendix:exp-details-node-classify}.

\subsubsection*{\bf Training and testing} Following the training protocol established in the BWGNN paper~\cite{GAnoDet-spectral-1-BWGNN}, we maintain a validation-to-test set split of $1:2$, and employ training ratios of $1\%$ (across all datasets) and $40\%$ (additionally for T-Finance). 
Baselines are trained for $100$ epochs using the Adam optimizer, without early stopping. 
We report the test results of the models that achieved the highest Macro-F1 score on the validation set, averaging results across $10$ random seeds to ensure robustness.


\section{Additional Results}
\label{appendix:additional-results}

In this section, we present additional results that bolster the experiments detailed in the main text, further substantiating our conclusions. 

\subsection{Full numerical experiment results}
\label{appendix:additional-results-full-numerical-exp}

We present a detailed overview of our numerical experiment results in Table~\ref{table-numerical-full}, including those for our TFGNN.

The data illustrates that both the trigonometric polynomial and TFGNN achieve outstanding performance, underscoring the advantages of our approach. Additionally, these results are consistent with the node classification outcomes outlined in Section~\ref{section-exp-node-classify}, validating the real-world applicability of our analysis.

\begin{table*}[!th]
\caption{Full numerical experiment results. introduced in Section~\ref{section-connect-polynomial-ability-spectral-GNN-ability-exp}. 
Both trigonometric polynomial and TFGNN are included for comprehensive evaluations.} 
\vskip -0.05in
\label{table-numerical-full}
\centering
\setlength{\tabcolsep}{4pt}
\resizebox{\textwidth}{!}{
\begin{tabular}{cc|cccccc|cccccc|cc}
\hline
\multicolumn{2}{c|}{Method}                         & \multicolumn{6}{c|}{Slice-wise approximation}                               & \multicolumn{6}{c|}{Filter Learning}                                        & \multirow{2}{*}{\makecell[c]{\# Avg \\ Rank 1}} & \multirow{2}{*}{\makecell[c]{\# Avg \\ Rank 2}} \\ \cline{1-14}
Polynomial    & \multicolumn{1}{c|}{GNN} & $f_{1}(x)$ & $f_{2}(x)$ & $f_{3}(x)$ & $f_{4}(x)$ & $f_{5}(x)$ & $f_{6}(x)$ & $f_{1}(x)$ & $f_{2}(x)$ & $f_{3}(x)$ & $f_{4}(x)$ & $f_{5}(x)$ & $f_{6}(x)$ &                              &                              \\ \hline
Monomial      & GPRGNN~\cite{GPRGNN}                              & 139.9      & 289.1      & 466.1      & 398.3      & 1.83       & 97.83      & 167.2      & 366.4      & 566.3      & 468.7      & 15.91      & 139.2      & 6                            & 6                            \\
Bernstein     & BernNet~\cite{BernNet-GNN-narrowbandresults-1}                             & 32.78      & 247.3      & 398.5      & 306.5      & 0.058      & 22.92      & 68.23      & 313.2      & 448.2      & 415.2      & 7.79       & 95.84      & 5                            & 5                            \\
Chebyshev     & ChebNetII~\cite{ChebNetII}                           & 23.45      & 85.19      & 244.8      & 187.2      & 0.018      & 13.13      & 64.22      & 168.4      & 402.5      & 347.5      & 6.83       & 86.25      & 4                            & 4                            \\
Jacobian      & JacobiConv~\cite{JacobiConv}                          & 22.18      & 80.77      & 239.2      & 155.3      & 0.017      & 11.82      & 48.56      & 95.92      & 338.1      & 266.4      & 5.33       & 65.13      & 3                            & 3                            \\
Learnable     & OptBasis~\cite{OptBasisGNN}                            & 20.75      & 80.53      & 225.7      & 152.7      & 0.017      & 11.20      & 43.44      & 89.48      & 289.5      & 238.1      & 4.98       & 61.70      & 2                            & 2                            \\ \hline
Trigonometric & TFGNN                               & \textbf{12.35}      & \textbf{23.69}      & \textbf{71.13}      & \textbf{59.88}      & \textbf{0.017}      & \textbf{6.52}       & \textbf{27.23}      & \textbf{65.19}      & \textbf{102.3}      & \textbf{105.3}      & \textbf{4.05}       & \textbf{48.08}      & \textbf{1}                            & \textbf{1}                            \\ \hline
\end{tabular}}
\end{table*}

\subsection{Additional ablation studies of $K$ and $\omega$}
\label{appendix:additional-results-ablation-k-omega}

\begin{figure*}[!t]
\centering
    \subfloat[Cora.]{
    \label{fig:cora-full}
    \includegraphics[width=0.23\linewidth]{figure/ablation_cora.pdf}
    }
    \subfloat[Citeseer.]{
    \label{fig:citeseer-full}
    \includegraphics[width=0.23\linewidth]{figure/ablation_citeseer.pdf}
    }
    \subfloat[Roman.]{
    \label{fig:roman-full}
    \includegraphics[width=0.23\linewidth]{figure/ablation_amazon.pdf}
    }
    \subfloat[Amazon.]{
    \label{fig:amazon-full}
    \includegraphics[width=0.23\linewidth]{figure/ablation_roman.pdf}
    } \\ 
    \subfloat[Pubmed]{
    \label{fig:pubmed-full}
    \includegraphics[width=0.23\linewidth]{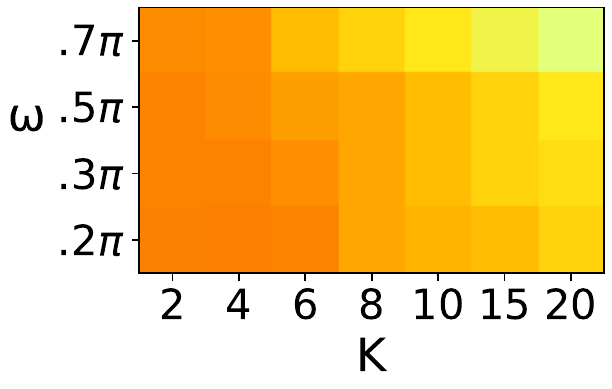}
    }
    \subfloat[Ques.]{
    \label{fig:question-full}
    \includegraphics[width=0.23\linewidth]{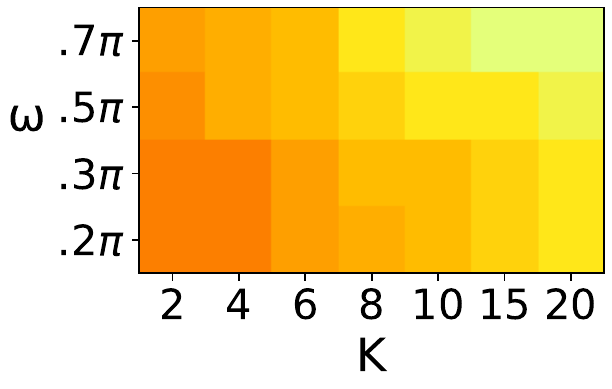}
    }
    \subfloat[Arxiv.]{
    \label{fig:arxiv-full}
    \includegraphics[width=0.23\linewidth]{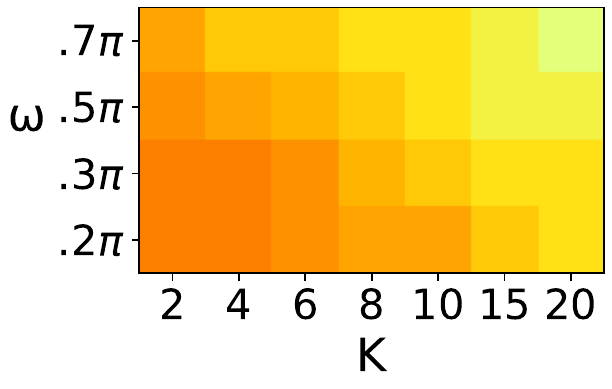}
    }
    \subfloat[Products.]{
    \label{fig:products-full}
    \includegraphics[width=0.23\linewidth]{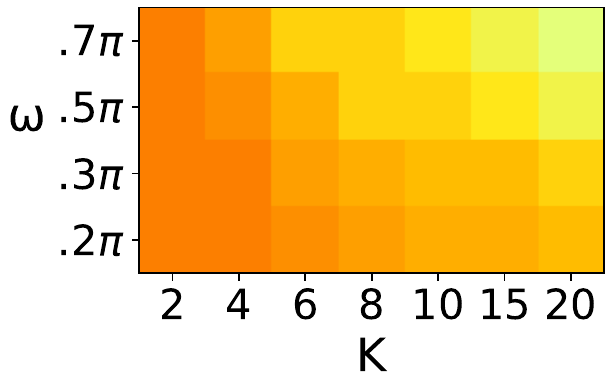}
    } \\
    \subfloat[Gamers]{
    \label{fig:gamers-full}
    \includegraphics[width=0.23\linewidth]{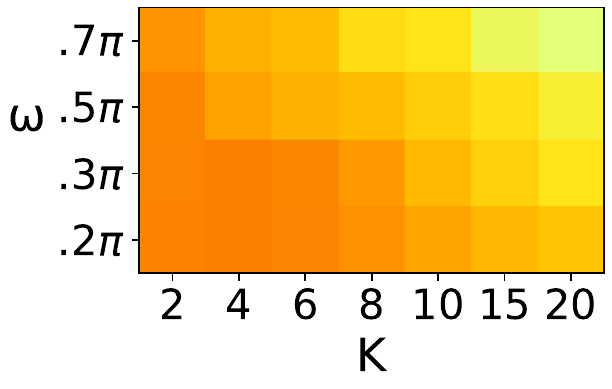}
    }
    \subfloat[Genius]{
    \label{fig:genius-full}
    \includegraphics[width=0.23\linewidth]{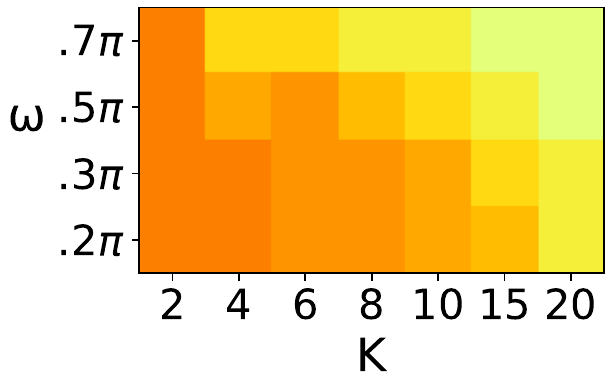}
    }
    \caption{Additional ablation studies on $K$ and $\omega$. 
    Darker shades indicate higher performance values.}
    \label{fig:ablation-K-omega-full}
\end{figure*}

In this section, we present an extended ablation study of the key hyperparameters $K$ and $\omega$, complementing our findings in Section~\ref{section-exp-node-classify-ablation}, with Figure~\ref{fig:ablation-K-omega-full} illustrating the outcomes.

The figures indicate a trend similar to  that highlighted in Section~\ref{section-exp-node-classify-ablation}, showing that the best-performing values for $K$, $\omega$, and the product $K\cdot\omega$ typically lie within low ranges.

\end{document}